\documentclass[11pt]{article}
\usepackage{amsfonts,enumerate,amssymb,amsmath}
\usepackage{epsfig}
\usepackage{subfigure,bm,cite}
\usepackage[applemac]{inputenc}
\usepackage{color}

\begin{document}

\title{Fractal Descriptors Based on Fourier Spectrum Applied to Texture Analysis}

\author{
{\bf Jo\~{a}o Batista Florindo, Odemir Martinez Bruno}\\
{\normalsize Instituto de F\'{i}sica de S\~{a}o Carlos - Universidade de S\~{a}o Paulo} \\ 
{\normalsize Av. Trabalhador S\~{a}o-carlense, 400 - CEP: 13560-970} \\ 
{\normalsize Cx. Postal 369 - S\~{a}o Carlos - SP - Brasil} \\ 
{\normalsize jbflorindo@ursa.ifsc.usp.br, bruno@ifsc.usp.br}  \\ \\
}
\date{}
\maketitle 
\thispagestyle{empty}

\noindent{}{\bf\large Abstract --} This work proposes the development and study of a novel technique for the generation of fractal descriptors used in texture analysis. The novel descriptors are obtained from a multiscale transform applied to the Fourier technique of fractal dimension calculus. The power spectrum of the Fourier transform of the image is plotted against the frequency in a log-log scale and a multiscale transform is applied to this curve. The obtained values are taken as the fractal descriptors of the image. The validation of the propose is performed by the use of the descriptors for the classification of a dataset of texture images whose real classes are previously known. The classification precision is compared to other fractal descriptors known in the literature. The results confirm the efficiency of the proposed method.
\\

\noindent{}{\bf\large Keywords --}
Fractal descriptors, fractal dimension, fractal multiscale, Fourier spectrum
\\

\section{Introduction}

Although the studies of fractal objects has more than a century, it is in the last decades that a growing number of works is presented by the literature, showing applications of fractals in several branches of the science. The fractals demonstrate ability in the representation overall of objects found in the nature, as already attested in \cite{M68}

Following this idea, one can find the work of Carlin \cite{C00} demonstrating the viability of a fractal dimension method in the calculus of fractal dimension of non fractal objects. Works like \cite{M68}, \cite{C00} and others encouraged the development of works describing natural objects by the fractal dimension \cite{PPFVOB05,CG90,G90}. These works experimented a lot of techniques for the extraction of image features based on fractal geometry and more specifically on methods for the calculus of fractal dimension, like Minkowski sausage \cite{T95}, Fourier \cite{R94}, etc.

Extending the concept of fractal dimension, \cite{MCSM02} presents the concept of multiscale fractal dimension, in which the fractal dimension is calculated at different scales of observation, allowing the obtainment of a set of numbers called the descriptors of the analyzed object, allowing for the enrichment of this sort of analysis. \cite{PPFVOB05} applied this concept, using the Minkowski sausage method, and obtained interesting results in the task of classification of Brazilian plants, based on the digitalized images of its leaves. \cite{BCB09} also applied the fractal descriptors to the analysis of texture images and also provided good results.

The present work proposes the development and study of a novel multiscale fractal descriptor for texture images. Here, the descriptors are obtained from the Fourier method for the fractal dimension calculus. In the Fourier technique, the fractal dimension is calculated from the power law detected in the graph of the power spectrum as a function of the frequency in the Fourier transform of the image. When this graph is plotted in a bi-log scale the curve is approximately similar to a straight line and the dimension is provided by the slope of such line.

The idea in this work is the application of a multiscale transform to the whole curve in log-log scale to provide the descriptors for the original image. In order to verify the power of the proposed technique in the description of texture images, the descriptors obtained are used in a process of classification of these images. The performance of the novel technique is validated by the comparison with other fractal descriptors known in the literature. We chose to compare only fractal analysis methods once the literature presents several works showing the efficience of fractal analysis face to conventional texture analysis technique in natural image analysis \cite{LDBBMB11,NC97,CDHLAB03,XYMLSR07,VGG98,PPFVOB05,BPFC08,BCB09}.

This work is divided into 8 sections. In the next, we show a brief study about fractal geometry and fractal geometry. The third section presents the Fourier fractal dimension. The fourth one describes fractal descriptors. In the fifth section, the proposed technique is described. The following section describes the experiments performed with the aim of validating the proposed technique. The seventh section shows the results. The eighth section expresses the conclusions of the work.

\section{Fractal Geometry}

Roughly speaking, a fractal is an object characterized by an infinite self-similarity and an infinite complexity. The self-similarity means that whether we observe a specific part of the object, this part behaves like a copy (or ``quasi-copy'') of a greater or smaller part of the same object and this process is repeated at any observation scale taken. In its turn, the concept of complexity lacks a precise definition but may be understood as the level of details observed at each different observation scale.

The most important measure used in the characterization of a fractal entity is its fractal dimension defined initially in \cite{M68}. This value estimates the spatial occupation of the fractal as well as its self-similarity degree.

For true fractals, \cite{M68} uses the Hausdorff-Besicovitch measure as the fractal dimension. The Hausdorff-Besicovitch dimension $dim_{H}$ of a metric space $X$ is given by:
\[
	dim_{H}(X) = inf{d \geq 0 | C_{H}^{d}(X) = 0},
\]
where $C_{H}^{d}(X)$ is the $d$-dimensional Hausdorff content of $X$, defined by:
\[
	C_{H}^{d}(X) = inf{\sum_{i}{r_{i}^{d} | \mbox{there exists a cover of x using balls of radii } r_{i} > 0}}
\]

Particularly, in the case of self-similar objects as the fractals, a simpler expression for the calculus of Hausdorff-Besicovitch dimension $D$, directly based on the calculus of the euclidean topological dimension and called similarity dimension, is given by:
\[
	D = \lim_{u\rightarrow 0}\frac{ln N}{ln L/u},
\]
where $N$ is the number of ``units'' $u$ needed to cover the whole extension $L$ of the object.

The above formula gave rise to the development of a lot of computational methods for the calculus of the fractal dimension of fractals represented as discrete objects in a digital image, which simplified significantly the obtainment of this measure. The methods used for the fractal dimension calculus in these cases may be essentially divided into two categories: the spatial methods, like box-counting \cite{CC00}, Bouligand-Minkowski \cite{T95}and mass-radius \cite{CC00} and the spectral methods, like Fourier \cite{R94} and wavelets \cite{JJ01}. Here, we focus on Fourier method for the calculus of the fractal dimension and this approach is described more minutely in the next section.

\section{Fourier Fractal Dimension}
\label{sec:Fourier}

The fractal dimension of an object can be calculated through the Fourier transform associated to a mass distribution $\mu(u)$, $u \in \Re^{n}$ and by using a classical Physics analogy. Let the Fourier transform of $\mu$ thus defined through:
\[
	\tilde{\mu(u)} = \int_{\Re^n}{e^{ix*u}d\mu(x)}.
\]
In the classical Physics, we have the definition of the s-potential $\phi_{s}$, $s \geq 0$ of the mass distribution $\mu$ by:
\[
	\phi_{s}(x) = \int{\frac{d\mu(y)}{|x-y|^{s}}}
\]
In the same way, the s-energy $I_{s}$ is obtained through:
\[
	I_{s}(\mu) = \int{\phi_{s}(x)d\mu(x)}.
\]
Applying these concepts to the mass distribution $\mu$:
\[
	\phi_{x}(x) = (|.|^{-s}*\mu) \equiv \int{|x-y|^{-s}d\mu(y)}.
\]
Performing the Fourier transform and using the convolution theorem:
\[
	\tilde{\phi}_{s}(x) = c|u|^{s-n}\tilde{\mu}(u).
\]
From the Parseval's theorem we thus have:
\[
	I_{s}(\mu) = (2\pi)^{n}c\int{\hat{\phi_{s}}(u)\overline{\tilde{\mu}(u)}du}.
\]
The energy is in this way easily obtained through:
\begin{equation}\label{eq:energy}
	I_{s}(\mu) = (2\pi)^{n}c\int{|u|^{s-n}|\hat{\mu}(u)|^{2}du}.
\end{equation}

The Hausdorff dimension of $C$ has its lower limit in $s$ if there is a mass distribution $\mu$ on the set $C \in \Re^2$ for which the expression \ref{eq:energy} is finite. Still, if $|\tilde{\mu}(u)| \leq b|u|^{-t/2}$, for a real constant $b$, then $I_{s}(\mu)$ always converges if $s < t$. The greatest $t$ for which there is a mass distribution $\mu$ on $C$ is called the Fourier fractal dimension of $C$.

Based on the mathematical theory, \cite{R94} proposes the use of the Fourier transform, more precisely, the power spectrum of the transform as a tool for the calculus of fractal dimension of texture images. \cite{R94} shows that the average Fourier power spectrum of the texture image obeys a power law scaling as described in \cite{M68}.

The power law is verified between the power spectrum $F_{h}$ and the frequency $f$ and is expressed by:
\[
	F_{h} \propto f^{2-\beta_{h}},
\]
being
\[
	\beta_{h} = 2.H_{n} + 2,
\]
where $H_{n}$ is the Hurst coefficient \cite{R94}.

\begin{figure}[htbp]
	\centering
	\begin{tabular}{ccc}
		\includegraphics[width=0.15\textwidth]{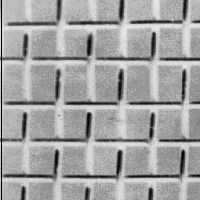} & \includegraphics[width=0.4\textwidth]{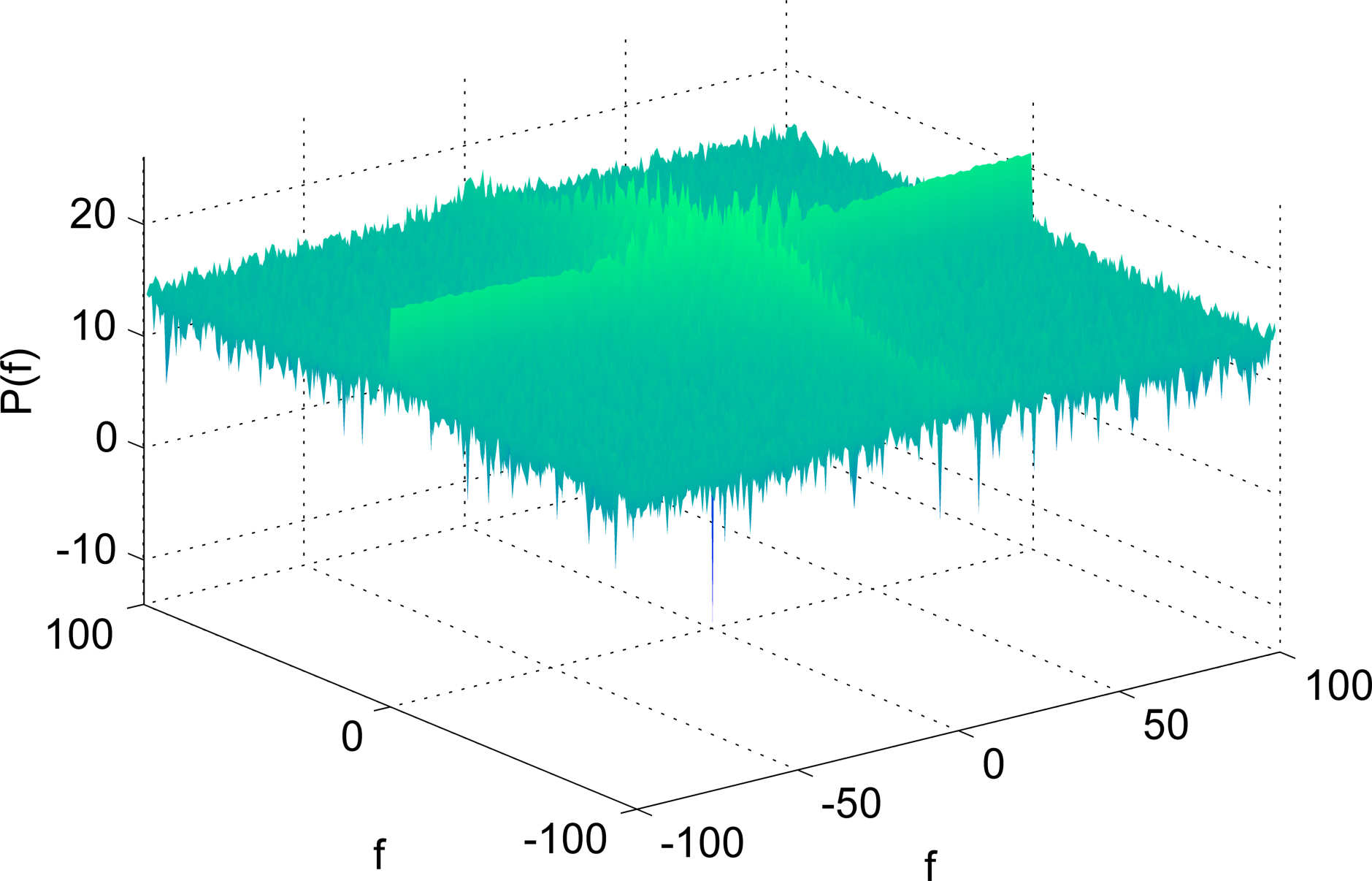} & 				   
		\includegraphics[width=0.4\textwidth]{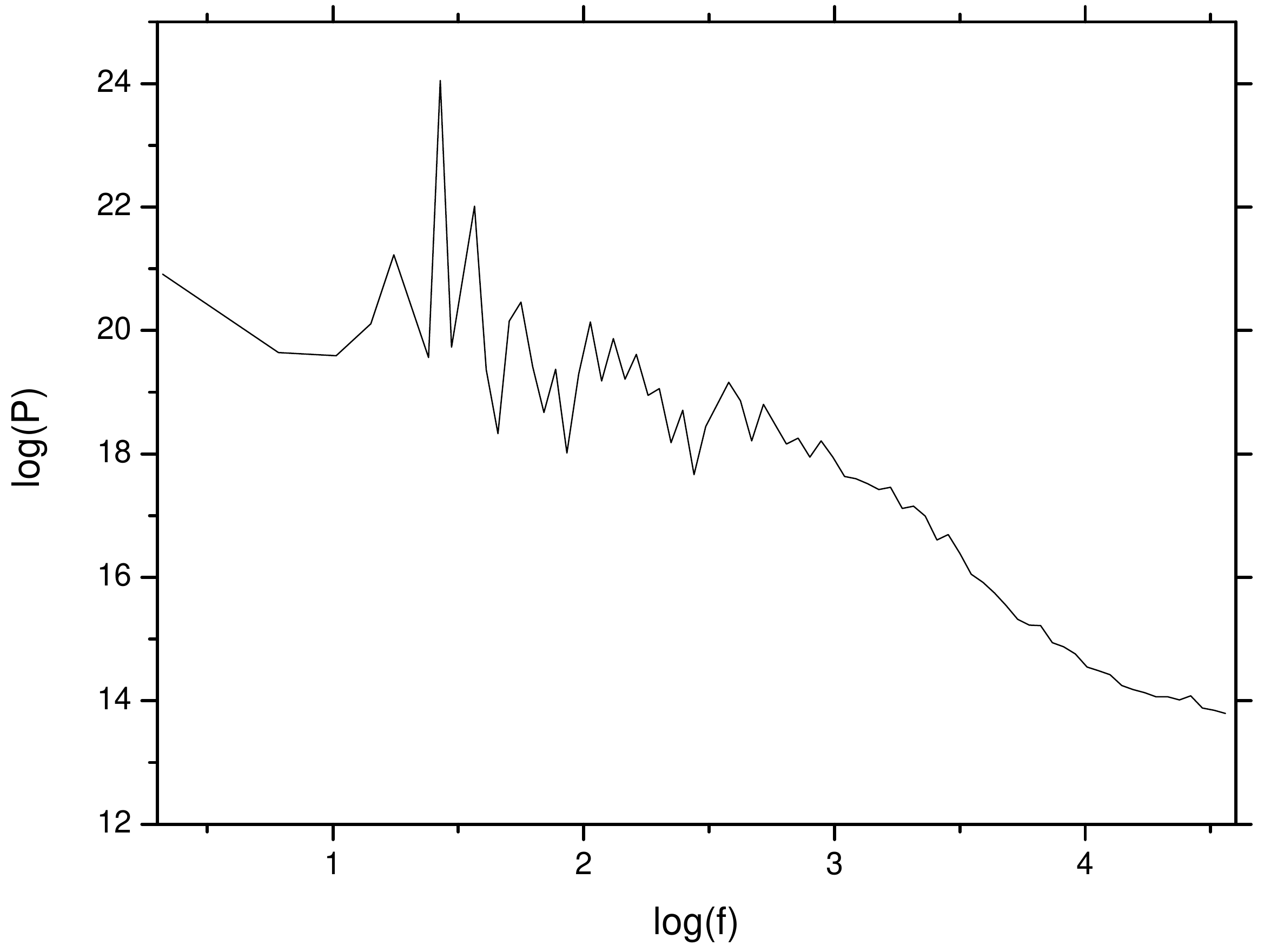}\\
	\end{tabular}
	\caption{Steps in fractal dimension calculus by Fourier method for texture images from two different classes. At left, the original texture. At center, the Fourier spectrum. At right, the plot of spectrum $\times$ frequency.}
	\label{fig:fourier}
\end{figure}

As a consequence of the power law, the plot in a log-log scale of the graph of $F_{h} x f$ results in a curve whose behavior approximates that of a straight line. \cite{R94} shows that the fractal dimension $FD_{h}$ of the image can be calculated by
\[
	FD_{h} = \frac{6+\beta_{h}}{2},
\]
where $\beta_{h}$ is the slope of the straight line best fitted to the curve in log-log scale. It is noticeable that the slope of this straight line at different directions is approximately constant, demonstrating the isotropism of the original image, allowing the application of Fourier fractal dimension calculus. The Figure \ref{fig:fourier} illustrates the process applied o 4 texture images (from 2 different classes).

\section{Fractal descriptors}

With the popularization of computer-aided image analysis techniques, it was observed also an increase in the amount of data processed by the computational algorithms. Such fact lead to the development of techniques with the aim of extracting relevant information in each image. The values encapsulating this essential information were called image descriptors and described important features from objects represented in the image.

Specifically, in texture analysis, a classical use of descriptors is the use of the Fourier descriptors \cite{GW02}, a reduced set of values describing the whole texture image which in its original format was meaningless for a classification algorithm, for example. The literature still showed the development of more robust descriptors, as Gabor \cite{GL09} and wavelets \cite{RBCDMMR95}, among others.

Mandelbrot \cite{M68} and other authors \cite{C00}\cite{R94} observed that many objects found in the nature present characteristics intrinsic to fractals, like the self-similarity and advanced levels of complexity. Such observations motivated authors like \cite{JJ01,G90,CG90} to employ the fractal dimension concept as a descriptor for objects of real world, represented in images.

Extending this idea, \cite{MCSM02} suggested the use of not only the fractal dimension as a unique descriptor but a set of values extracted from the fractal dimension measure process to characterize an image. Finally, \cite{BPFC08} formalized the concept of fractal descriptors as being a set of values extracted from fractal geometry methods and used to characterize artifacts in an image, like textures, contours, shapes, etc. We may still see examples of applications of fractal descriptors in \cite{PPFVOB05,BCB09,FCB10,Florindo2011ijmpc,Florindo2012prl,Florindo2011chaos}, among others.

Particularly, in this work, we present a novel method for fractal descriptors based on the fractal dimension calculated by using the Fourier spectrum of the image, described in the Section \ref{sec:Fourier}.

\section{Proposed Method}

Following the idea of the aforementioned works, here we propose the development of a novel fractal descriptor for texture images. We explore the obtainment of a set of descriptors by applying a multiscale approach to the technique of fractal dimension calculus by Fourier transform. Thus, we initially define the concept of multiscale transform.

\subsection{Multiscale Transform}

In practical situations, often we are faced with the need to analyze signals or images presenting different structures and patterns at different observation scales. In order to capture such scale variance, \cite{W84} proposed the development of multiscale transform. Still a lot of works \cite{MCSM02,WM08,AMMNPPST07} applied the multiscale transform for the analysis of experimental data in different situations.

Essentially, the multiscale transform of a signal $u(t)$ is the function $U(b,a)$, where $b$ is directly associated with $t$ and $a$ is the scale variable. Roughly speaking, the multiscale transform may be accomplished through three approaches: scale-space, time-frequency, time-scale. Each approach is briefly described in the following. More details may be seen in \cite{CC00}.

\subsubsection{Scale-space}

Scale-space transform is the most used multiscale transform. It is based on gaussian kernel properties \cite{W84} and may be expressed by the following expression:
\[
	\{(b,a)|a,b \in \Re, a > 0, b \in \{U^1(t,a)\}_zc\},
\]
where $._zc$ denotes for the zero-crossings of the expression $.$ and $U^1(t,a)$ expresses the convolution of the original signal $u(t)$ with the first derivative of the gaussian $g^1_a$, that is:
\[
	U^1(t,a) = u(t) * g_a^1(t).
\]

\subsubsection{Time-frequency}

\cite{W84} demonstrates that the scale $a$ and the frequency $f$ are simply related through $a = 1/f$. This result allows for the use of frequency transforms in multiscale approaches. Specifically, the time-frequency transform is based on the short-time Fourier transform, using a gaussian kernel. The following expression summarizes the transform:
\[
	U(b,f) = \int_{-\infty}^{+\infty}g*(t-b)u(t)e^{-i2\pi ft}dt,
\]
where $g*$ is the conjugate of the classical gaussian. In practice, the transform is performed through Gabor filters \cite{CC00}.

\subsubsection{Time-scale}

Finally, we have the time-scale transform, in which the slide window used in the short-time Fourier transform for time-frequency transform has variable side length allowing a better precision in the frequency localization. In this context, we notice that wavelet transform \cite{CC00} is the best candidate for the calculus of such particular transform. Thus, the time-scale transform may be represented through:
\[
	U\Phi(b,a) = \frac{1}{\sqrt(a)}\int_{R}\Phi*(\frac{t-b}{a})u(t)dt,
\]
where $\Phi$ is the classical mother wavelet and $*$ denotes its conjugate.

\subsection{Extracting Information from the Multiscale Transform}

A particular aspect of multiscale transform is that it maps a one-dimensional signal onto a two-dimensional function, resulting in information redundancy. In order to solve such drawback the literature showed some proposes. For instance, \cite{M92} proposed the use of local maxima, minima and zeros of the transform $U(b,a)$. \cite{RV93} used the projection of $U(b,a)$ onto the scale (equivalent to frequency) axis. Finally, \cite{B90} presented the use of a specific scale $a_0$.

A particular aspect of multiscale transform is that it maps a one-dimensional signal onto a two-dimensional function, resulting in information redundancy. In order to solve such drawback the literature showed some proposes. For instance, \cite{M92} proposed the use of local maxima, minima and zeros of the transform $U(b,a)$. \cite{RV93} used the projection of $U(b,a)$ onto the scale (equivalent to frequency) axis. Finally, \cite{B90} presented the use of a specific scale $a_0$.

In this work, we opted for the analysis of the multiscale transform of Fourier fractal dimension through the last approach, selecting a specific scale. Such chosen was motivated by the fact that we cannot priorize any region of log-log curve in our application, once the fractal characteristic is sensibly altered if we do not take into account the whole curve.

As described in details in the Section \ref{sec:Fourier}, the Fourier fractal dimension is calculated by the slope of a straight line fitted to the curve of power spectrum $\times$ frequency in the log-log scale. The method used to calculate the Fourier fractal dimension presents an inherent multiscale characteristic given by the use of the frequency as a scale variable. As demonstrated in \cite{W84}, in a multiscale analysis, frequency and scale are two ways of expressing the same property. In fact, by the Fourier filtering theory \cite{CC00}, we know that the lower frequencies express a general ``view'' of the image while the higher express more detailed structures in the original image. In order to emphasize this multiscale aspect, we propose the application of a multiscale transform to the curve in log-log scale.

In its turn, the inherent constitution of image textures implies that these textures contain relevant information stored in specific patterns of fine and coarse structures (more and less detailed). This is a strong motivation for the use of the fractal dimension being an interesting tool to describe such texture, quantifying the degree of complexity (details) at different scales of observation.

In addition to this, the use of the multiscale transform applied to the fractal dimension allows for a richer description of this complexity pattern of the image. Instead of use the unique value of fractal dimension, with the multiscale transform, we obtain a curve with a set of values, which represents in a more faithful manner the original texture, by expressing not only the complexity of texture as a whole but measuring the complexity at each different scale.

Based on this assumption, \cite{MCSM02} studied the extraction of multiscale features from the fractal dimension and used such features in a genic expression problem, obtaining interesting results. \cite{PPFVOB05} and \cite{BPFC08} applied the scale-space transform to the Bouligand-Minkowski fractal dimension and used the generated values as descriptors in a shape recognition task. \cite{BCB09} used the multiscale transform of Bouligand-Minkowski in a texture recognition problem.

\begin{figure}[htbp]
	\centering
	\begin{tabular}{c}
		\includegraphics[width=0.7\textwidth]{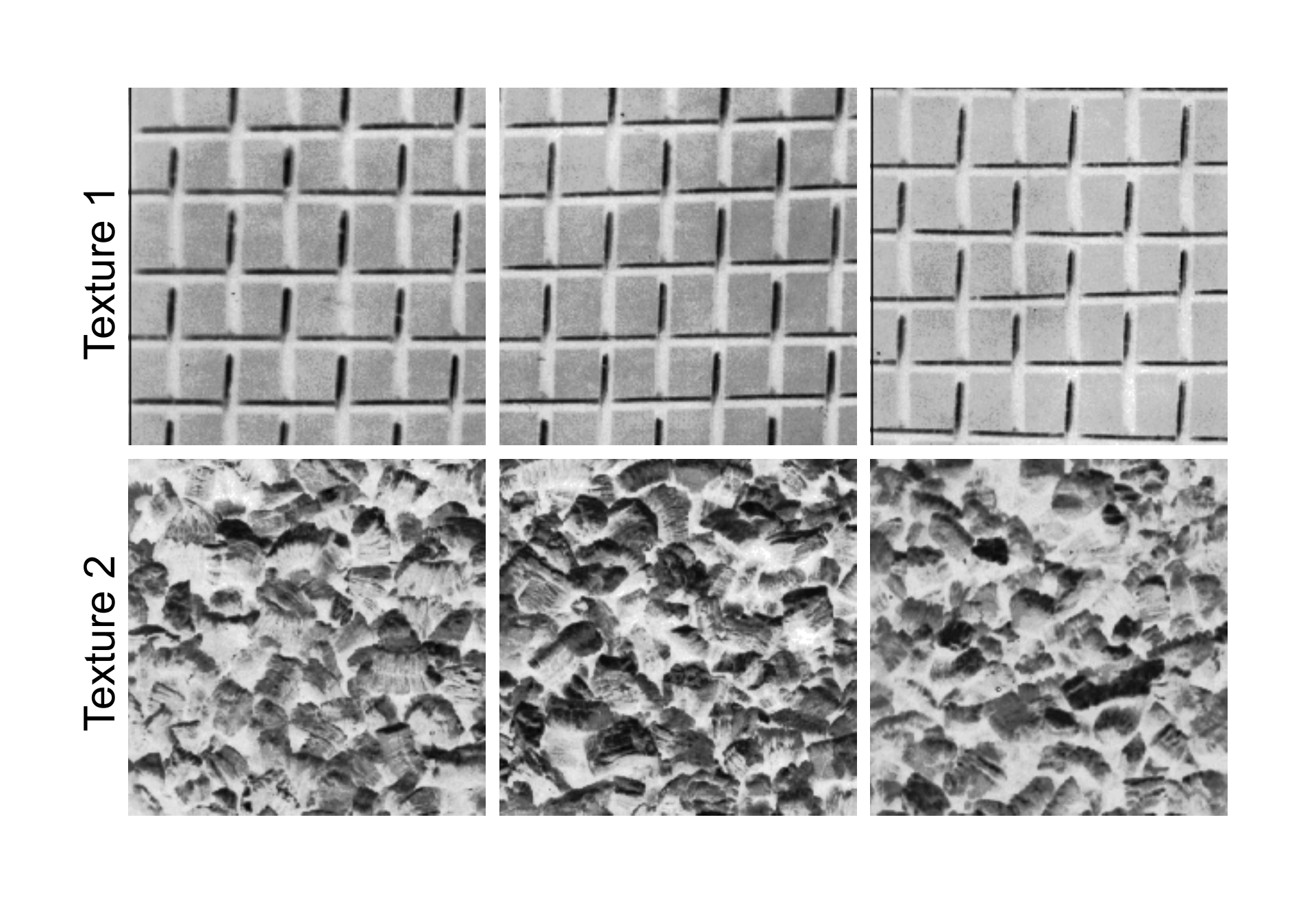}\\
		\includegraphics[width=0.7\textwidth]{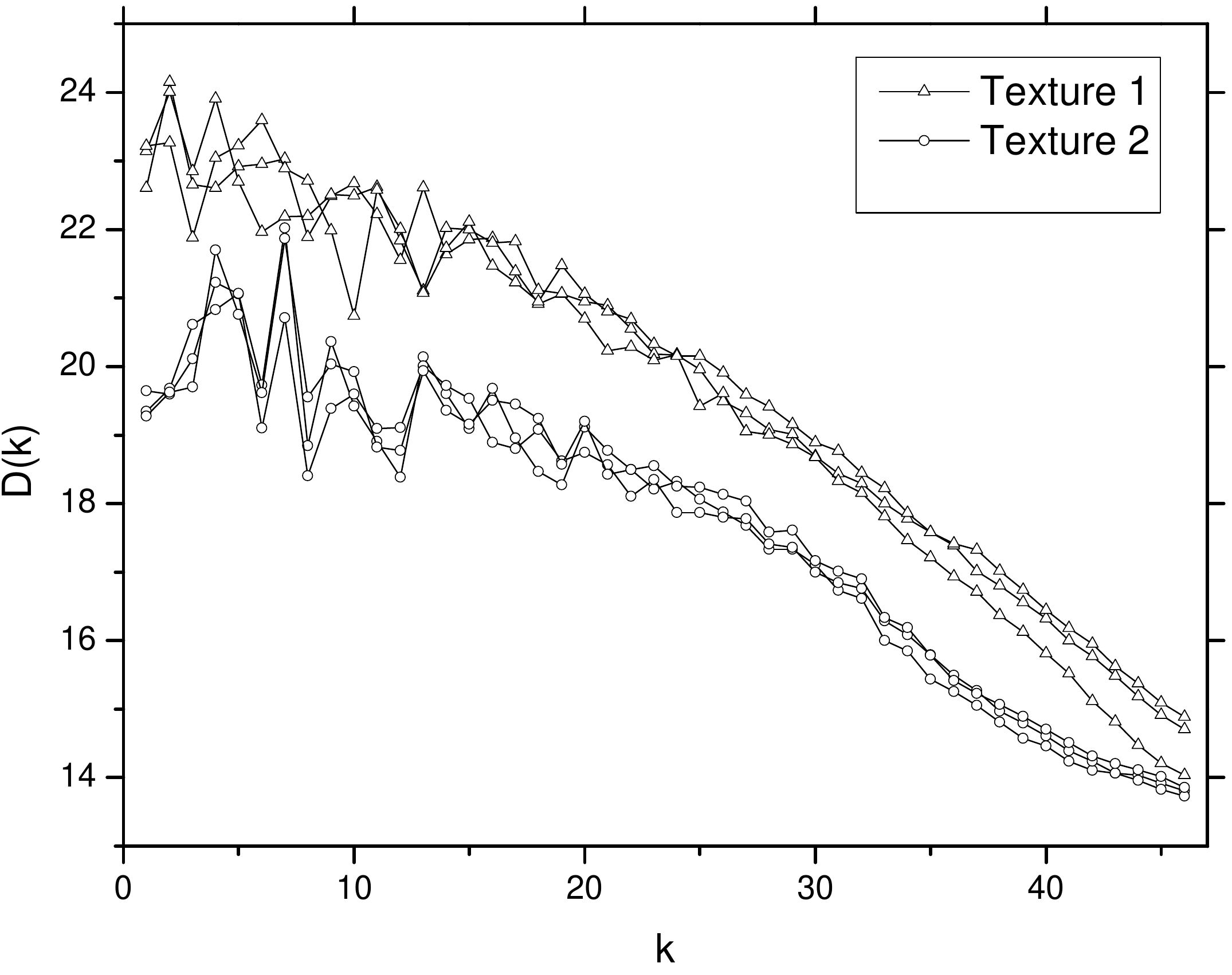}\\
	\end{tabular}
	\caption{Fourier fractal descriptors $D(k)$ of textures extracted from different classes.}
	\label{fig:classes}
\end{figure}

All these works applying multiscale to the fractal dimension presented in the literature demonstrated the accuracy of this approach in image analysis, presenting advantage over the classical methods used for this purpose. With this in mind, here we propose a novel multiscale fractal approach, comparing it with other known fractal descriptors approaches. For the generation of descriptors, we propose the application of a multiscale transform over the curve of $\log(power spectrum) \times \log(frequency)$ described in Section \ref{sec:Fourier}. Based on empirical verifications we opted for the use of scale-space transform. The Figure \ref{fig:classes} exemplifies like such descriptors can differentiate textures extracted from two different scenes (classes).

\section{Experiments}

The validation of the proposed technique is implemented by means of a classification experiment of datasets of texture images whose classes are predetermined.

The experiments are executed over four texture datasets. The first is the classical texture dataset of Brodatz \cite{B66}, complete with 10 windows extracted from each texture sample, constituting in this way 111 classes with 10 samples in each class. The second dataset is composed by textures of real scenaries and is named USPTex \cite{}. This dataset is composed by 191 classes with 12 images in each class. The third one is the OuTex dataset \cite{OMPVKH02} for tests. This is composed by 68 classes with 20 natural texture images in each class. Finally, we used a plant leaves data set \cite{BCB09}, composed by images from the leaf texture of different brazilian plant species. This data consists in 10 classes with 200 samples in each class.

The classification properly is executed by the Bayesian classifier technique \cite{DH00} implemented in weka environment \cite{H08}. The Bayesian method was chosen since, although it obtains lower correctness rates in the classification process, it corresponds to an optimal classifier and is less susceptible to particular characteristics of each compared descriptor or dataset. The accuracy of the classification is compared to other descriptor methods found in the literature and based on the fractal theory, that is, fractal Brownian motion \cite{CCCTC98}, box-counting \cite{GB08}, Bouligand-Minkowski volumetric fractal descriptors \cite{BCB09} and multifractal probability spectrum \cite{H01}. We compare the global correctness rate of each descriptor and the behavior of each method varying the number of descriptors.

We opted to compare only fractal methods due to the scope of this work which emphasizes texture fractal analysis. Moreover, the literature shows a lot of works comparing and demonstrating the power of fractal analysis face of classical texture analysis methods. The interested reader may, for instance, consultate such comparison in works like \cite{LDBBMB11,NC97,CDHLAB03,XYMLSR07,VGG98,PPFVOB05,BPFC08,BCB09}.

Relative to the number of descriptors an important aspect to be stated is that in fractal approach we do not limit the number of descriptors according to the number of samples per class. This procedure suggested in works like \cite{KDM00} is fundamental when descritptors are a set of features. Nevertheless, in fractal analysis, the set of descriptors must be thought as a unique correlated feature with a multiple dimension. Even the results in the next section comprobes that the use of a reduced set of fractal descriptors compromises seriusly the performance of the method.

\section{Results}

\begin{figure}[!htpb]
\centering
\includegraphics[width=0.5\textwidth]{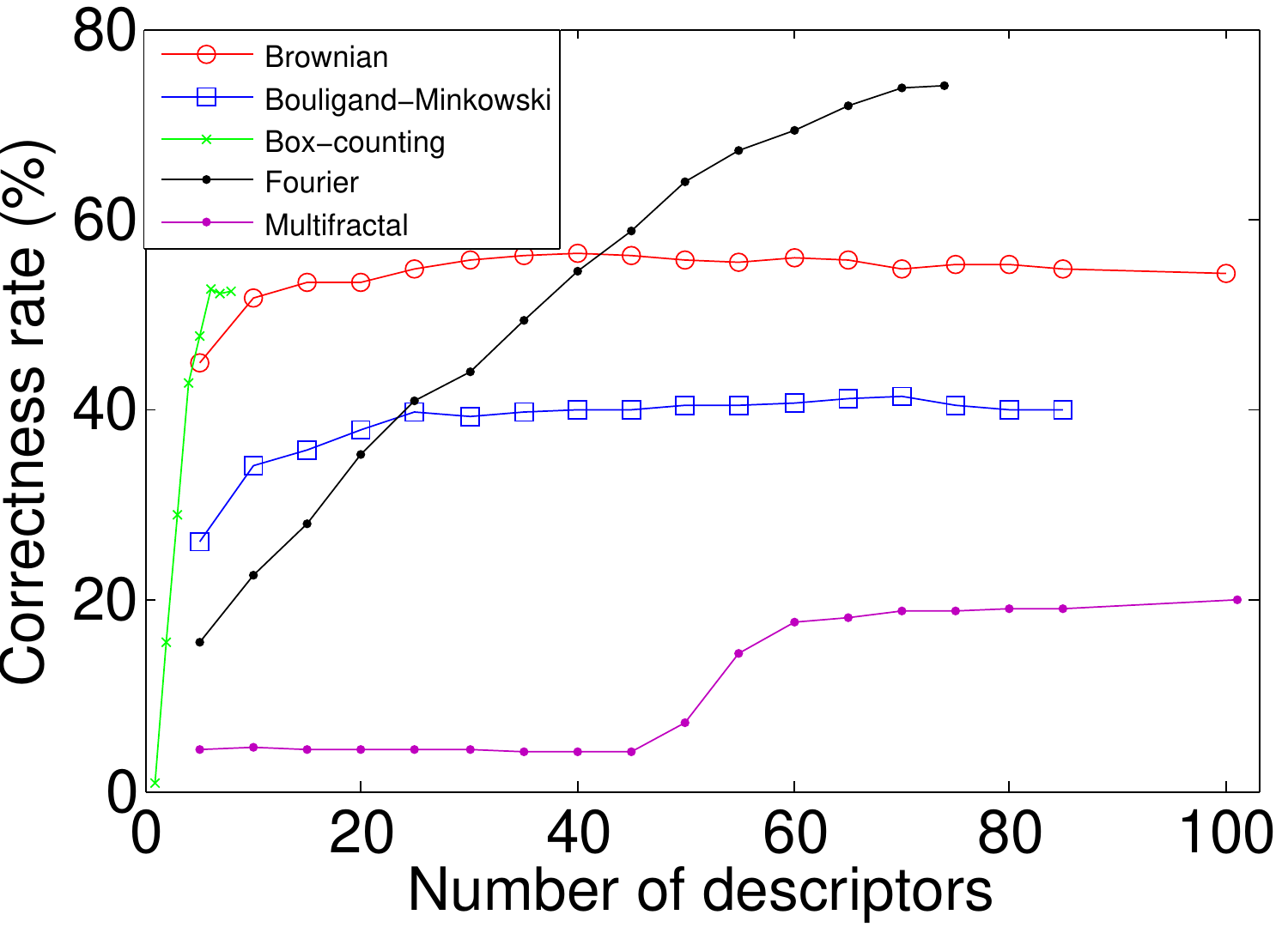}
\caption{Correctness rate for each descriptor method and number of descriptors in Brodatz dataset.}
\label{fig:brodatz_desc}
\end{figure}

The Figure \ref{fig:brodatz_desc} exhibits the correctness rate (percentage of images correctly classified) when varying the number of descriptors used in the Brodatz dataset. We observe that Bouligand-Minkowski and Brownian have a similar global aspect, presenting a huge variation with few descriptors and tending to stabilize when the number of descriptors incresases. Such behavior is justifiable by the nature of such methods, in which the first descriptors capture more faithfully the nuances in the object, in this case, the surface representing the texture image. In the box-counting experiment, the number of descriptors is limited by $log_2^N$, where $N$ is the side length of the image, once for each descriptor the side length of the box is doubled. The correctness rate also presents an exponential growing as the number of descriptors grows. Finally, the proposed technique presents a relation approximately linear of the correctness relative to the number of descriptors. Such appearance of the curve was waited, due to the linear nature of the Fourier transform.

\begin{figure}[!htpb]
\centering
\includegraphics[width=0.5\textwidth]{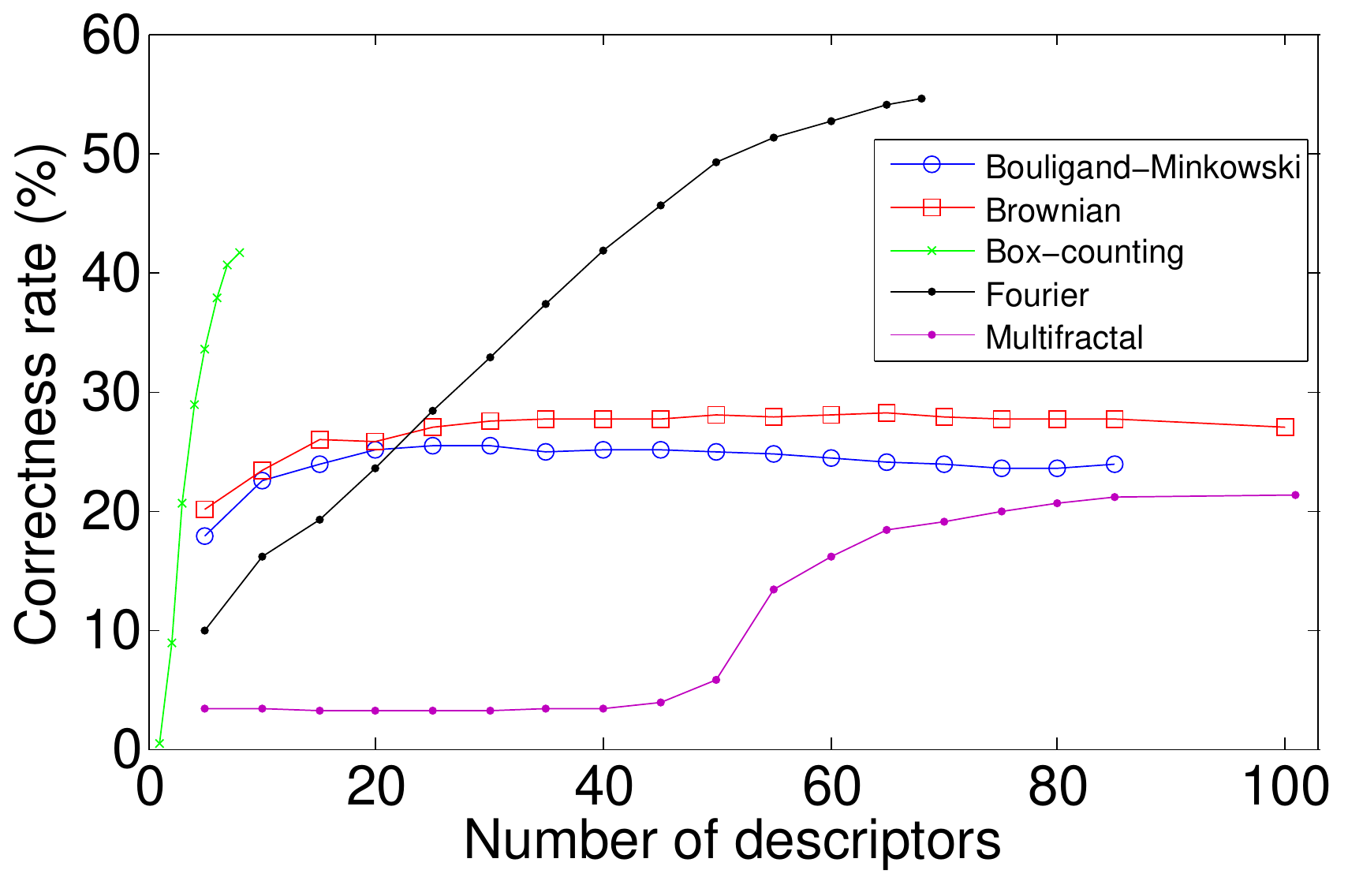}
\caption{Correctness rate for each descriptor method and number of descriptors in USPTex dataset.}
\label{fig:usptex_desc}
\end{figure}

\begin{figure}[!htpb]
\centering
\includegraphics[width=0.5\textwidth]{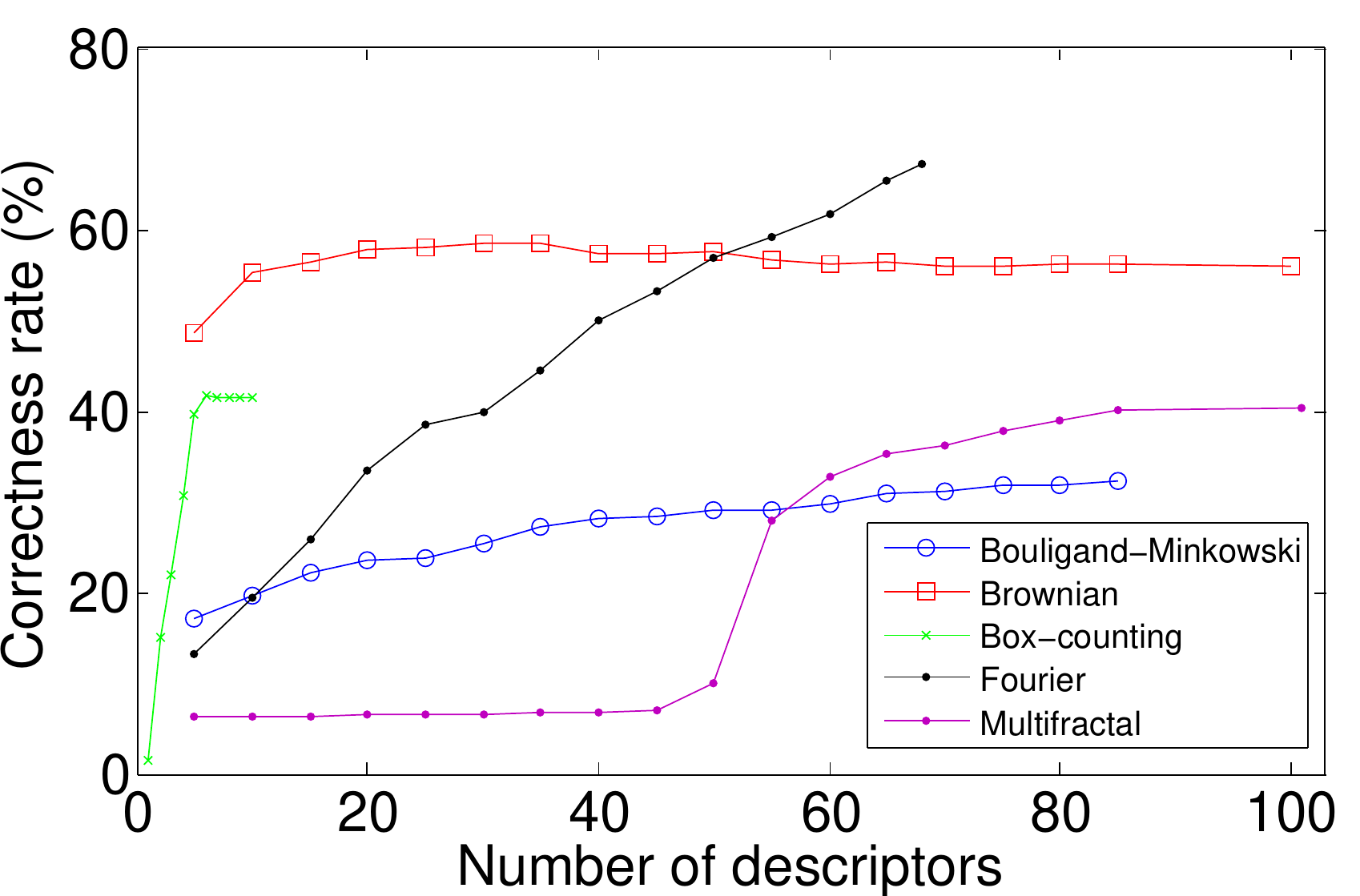}
\caption{Correctness rate for each descriptor method and number of descriptors in OuTex dataset.}
\label{fig:outex_desc}
\end{figure}

It is important to point out that although other methods present best results than Fourier when using few descriptors, the correctness tends to stabilize from a certain number of descriptors. This fact is explained by the so-called curse of dimensionality.
In its turn, the Figure \ref{fig:usptex_desc} shows the curves of correctness rate relative to the number of descriptors in the USPTex dataset. The global aspect of curves is similar to that seen in the Figure \ref{fig:brodatz_desc}. Nevertheless, the correctness rate as a whole is significantly smaller in this dataset. This decrease is explained by the greater number of classes present in USPTex dataset, embarrassing the discrimination among so heterogeneous samples.
The Figure \ref{fig:outex_desc} depicts the correctness rate for the compared methods, when we vary the number of descriptors in the OuTex dataset. The result is similar to that of the other datasets and the best result is achieved through Fourier method.

\begin{figure}[!htpb]
\centering
\includegraphics[width=0.5\textwidth]{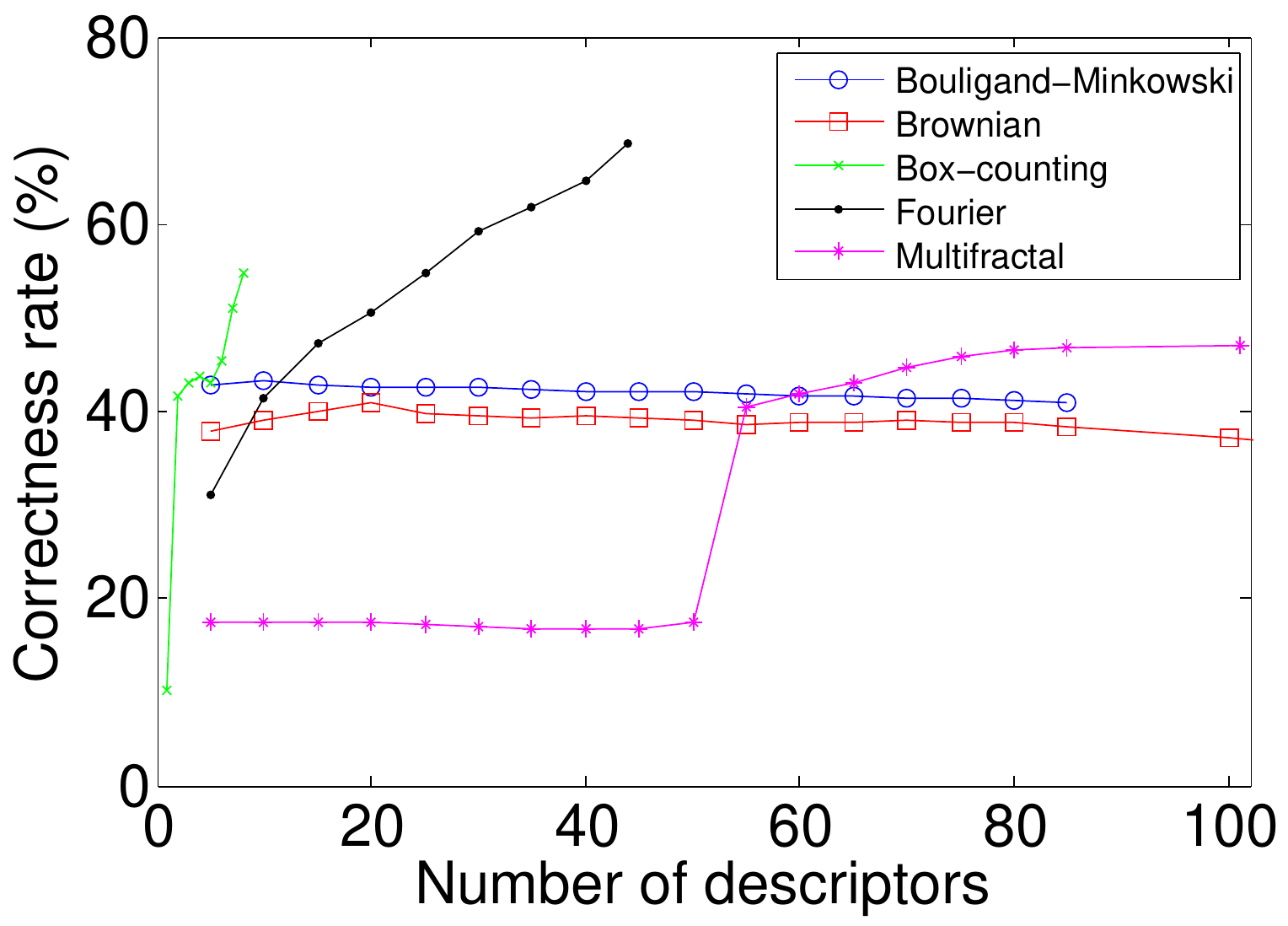}
\caption{Correctness rate for each descriptor method and number of descriptors in the plant leaves dataset.}
\label{fig:folhas_desc}
\end{figure}
\begin{figure}[!htpb] 
					 \centering
           \mbox{\subfigure[]{\includegraphics[width=7cm]{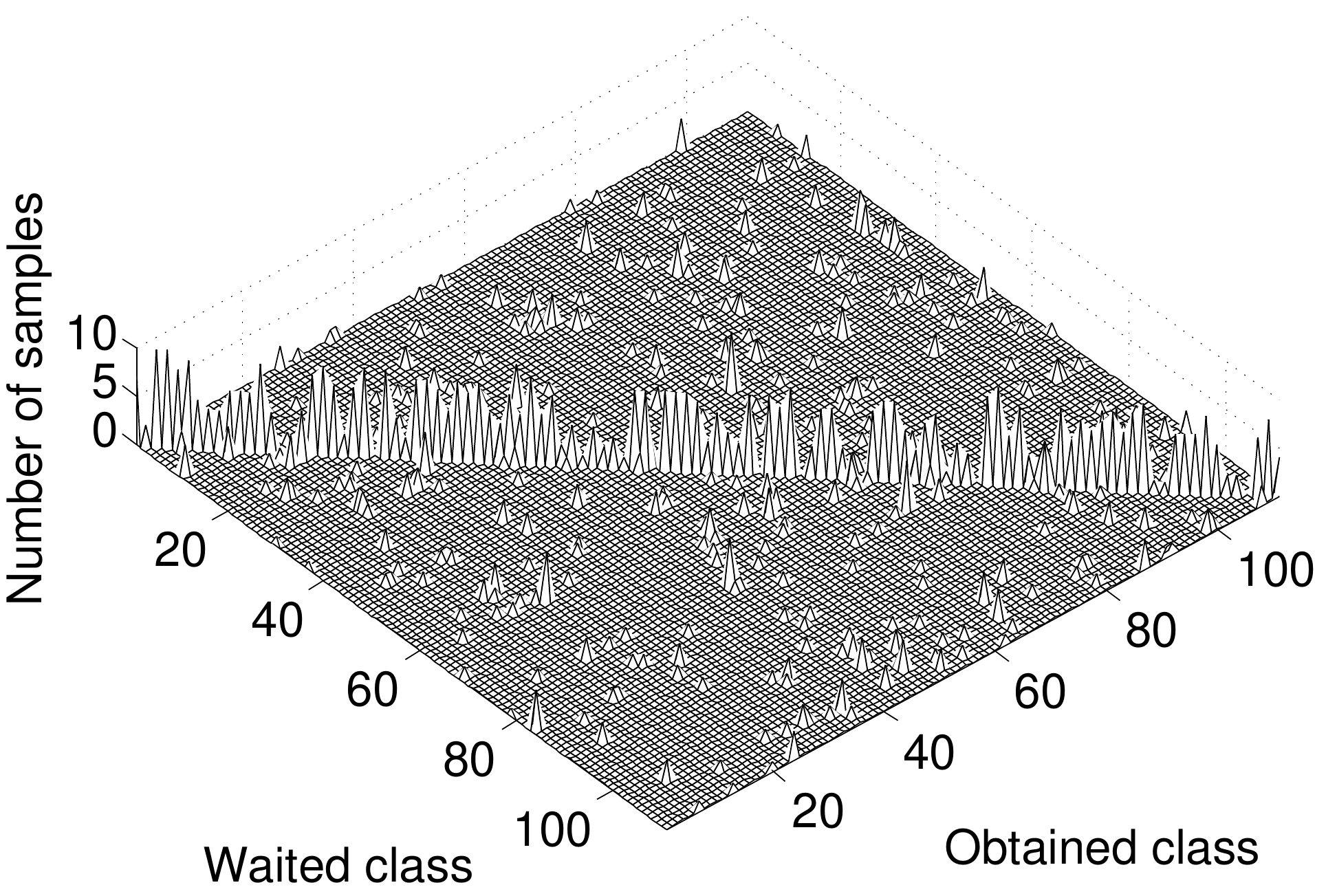}}
           			 \subfigure[]{\includegraphics[width=7cm]{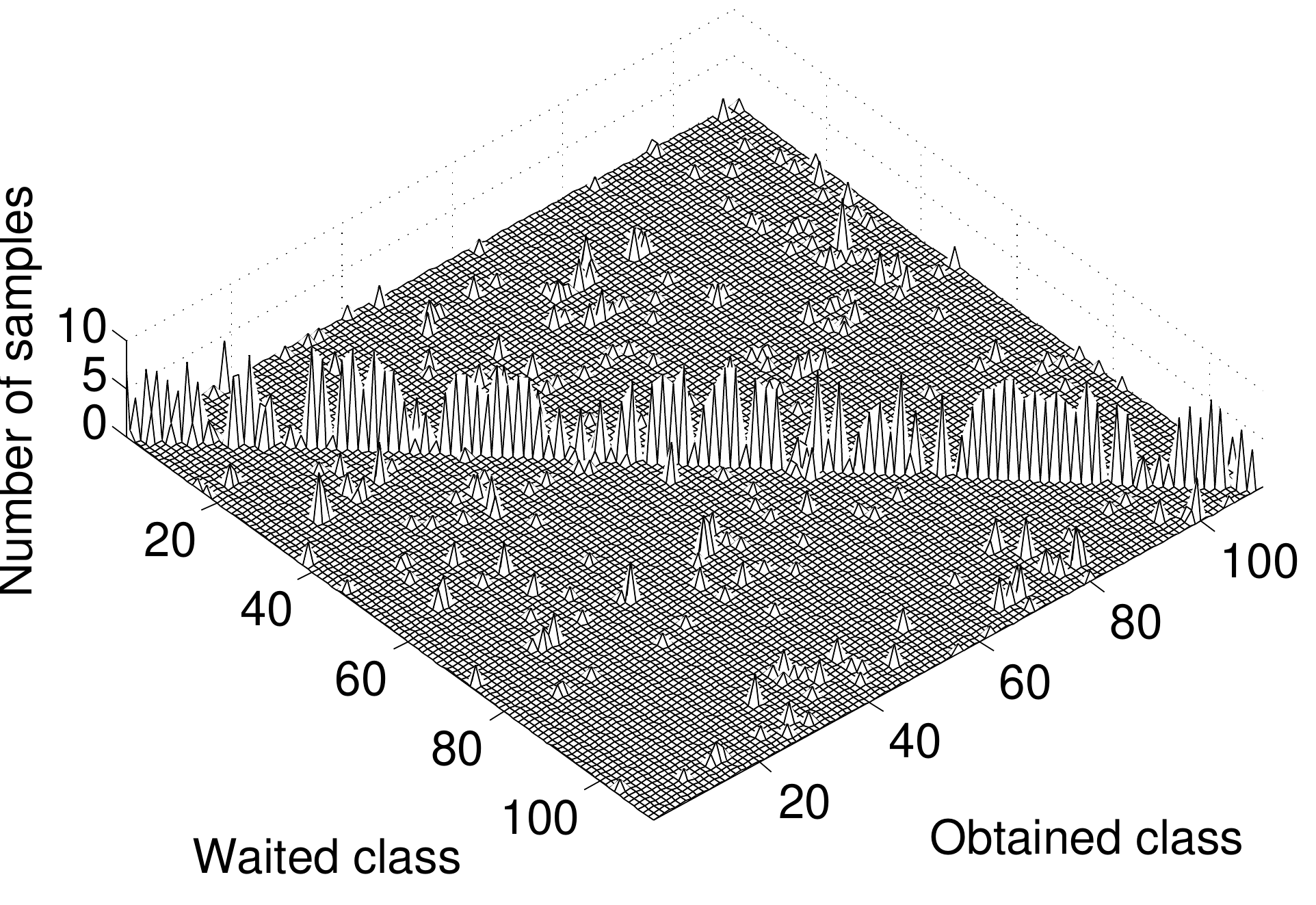}}}
					 \mbox{\subfigure[]{\includegraphics[width=7cm]{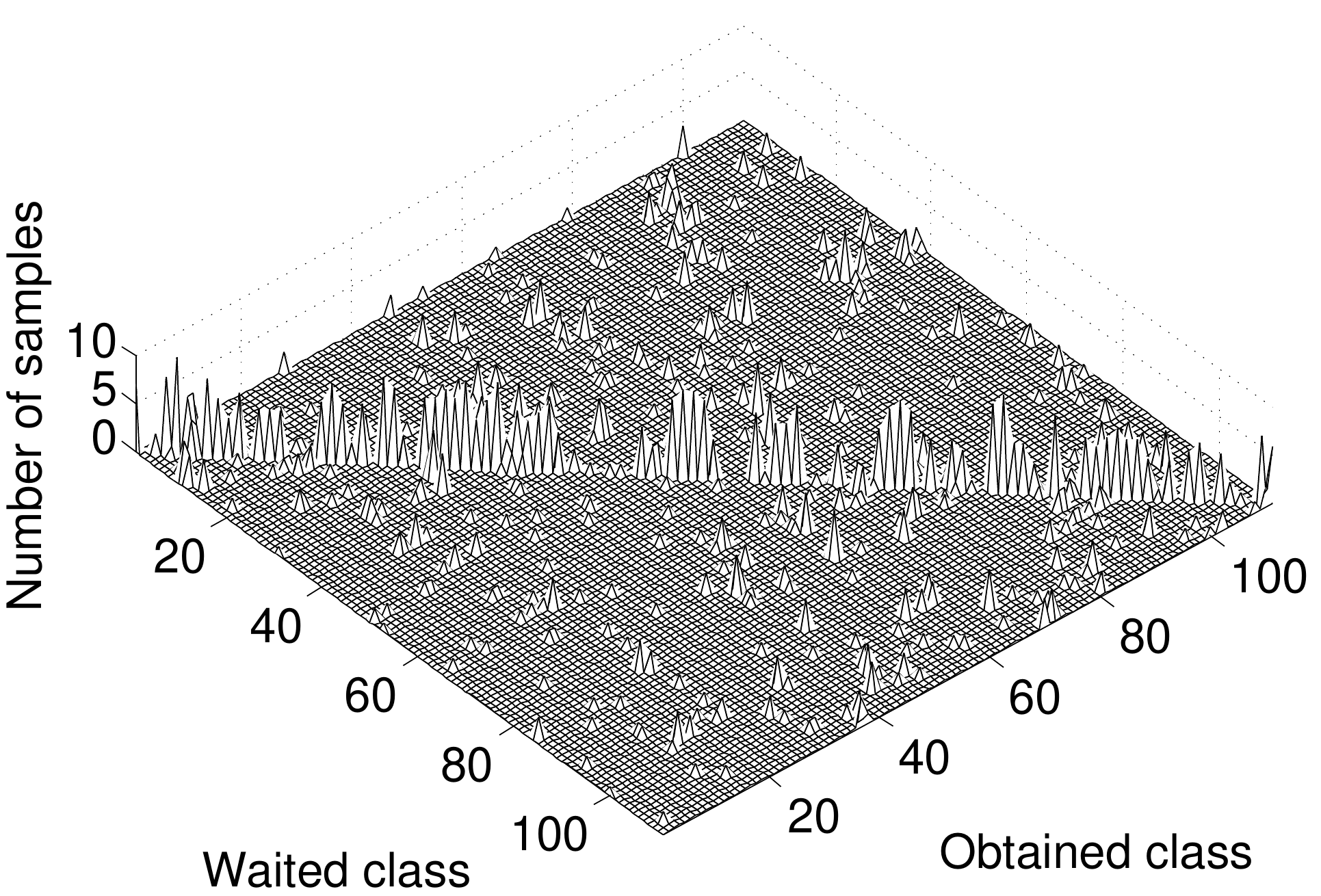}}
								 \subfigure[]{\includegraphics[width=7cm]{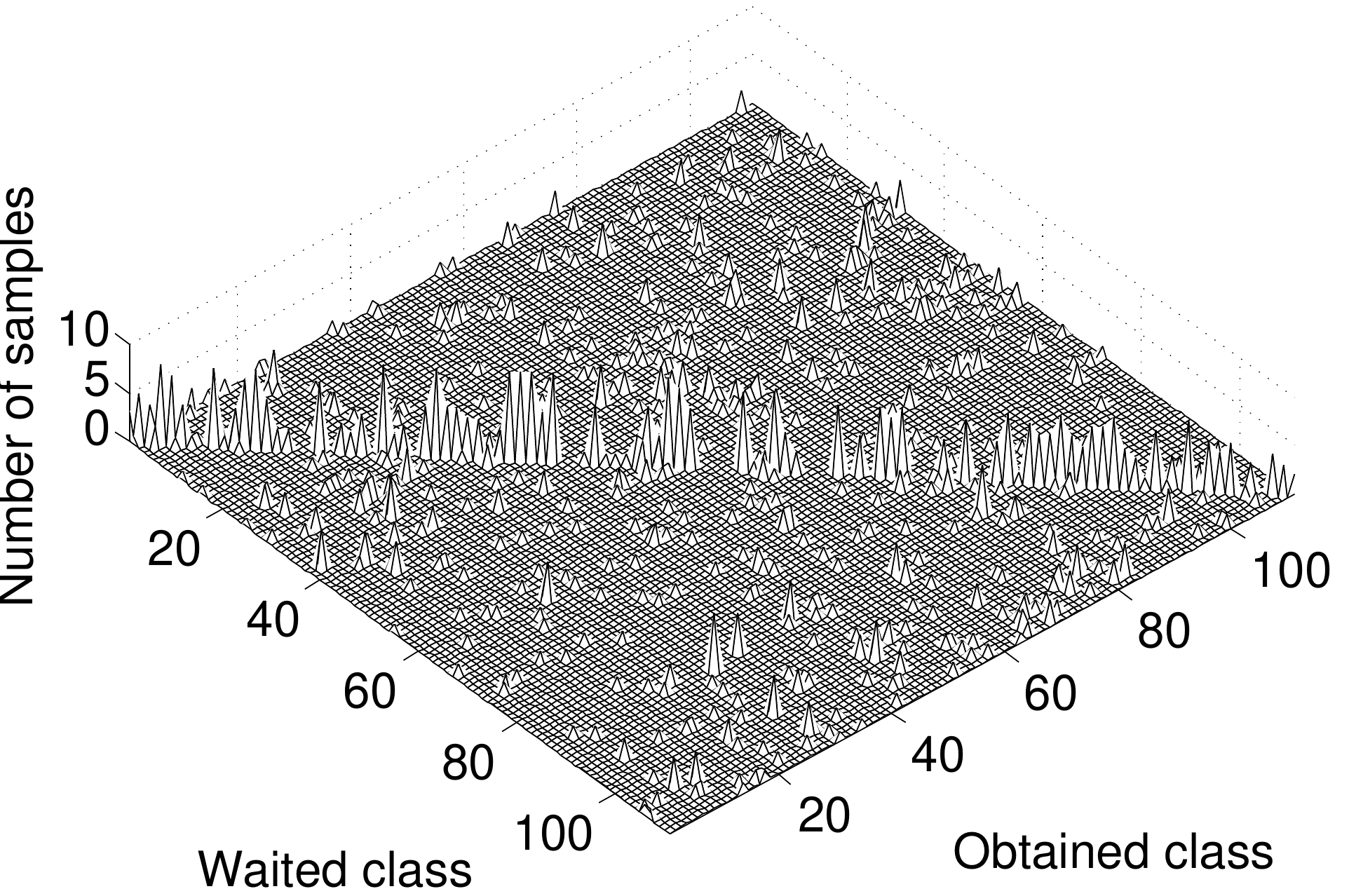}}}					 
					 \mbox{\subfigure[]{\includegraphics[width=7cm]{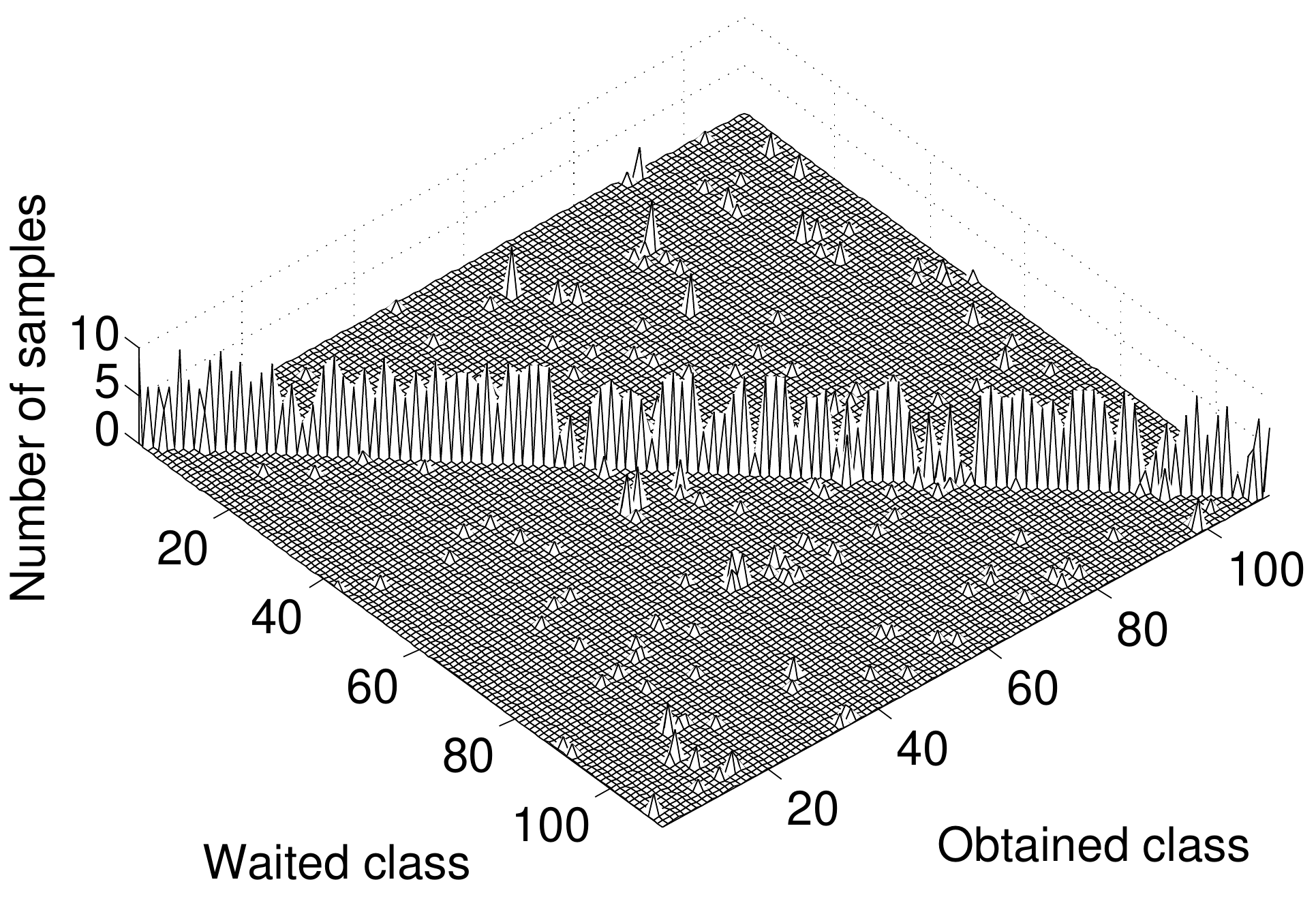}}}								 
           \caption{Confusion matrices for each descriptor method in Brodatz dataset. Each matrix is mapped onto an image in wich each value in the original image corresponds to a color in the image (see the colorbar legend). a) Box-counting. b) Brownian. c) Bouligand-Minkowski. d) Multifractal. e) Fourier.}
           \label{fig:brodatz_conf}                                  
   \end{figure}
\begin{figure}[!htbp] 
					 \centering
	 \mbox{\subfigure[]{\includegraphics[width=7cm]{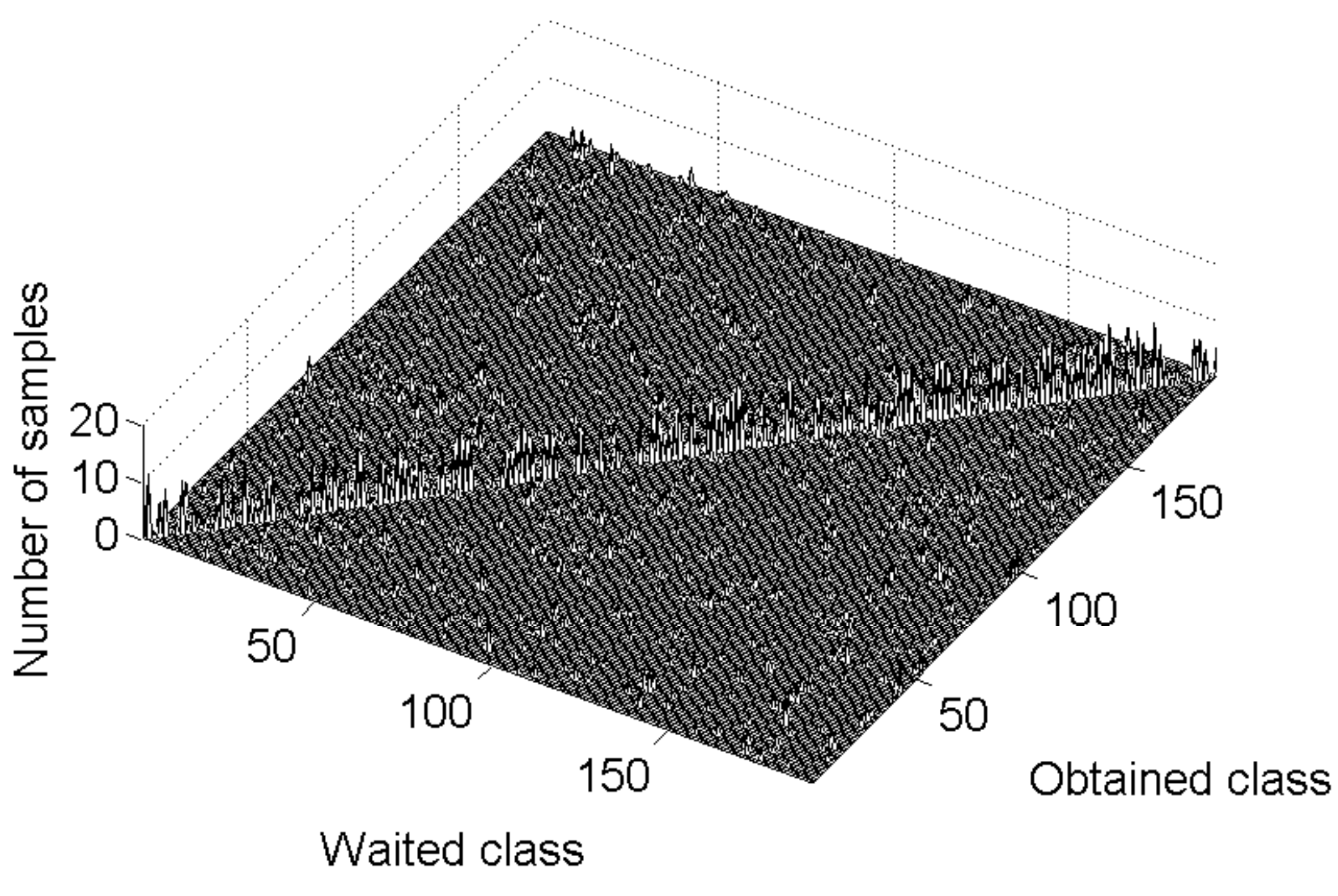}}
           			 \subfigure[]{\includegraphics[width=7cm]{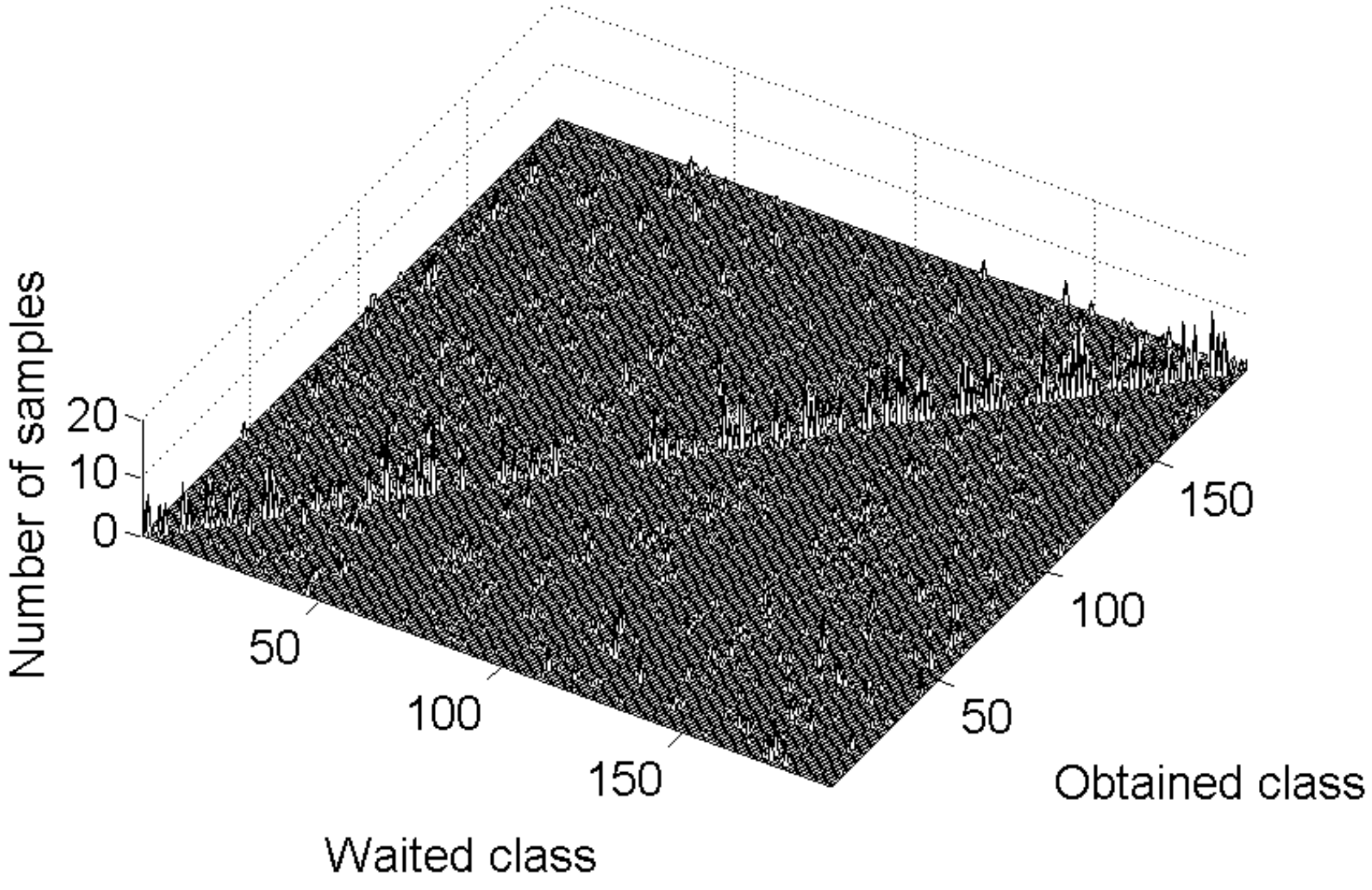}}}
					 \mbox{\subfigure[]{\includegraphics[width=7cm]{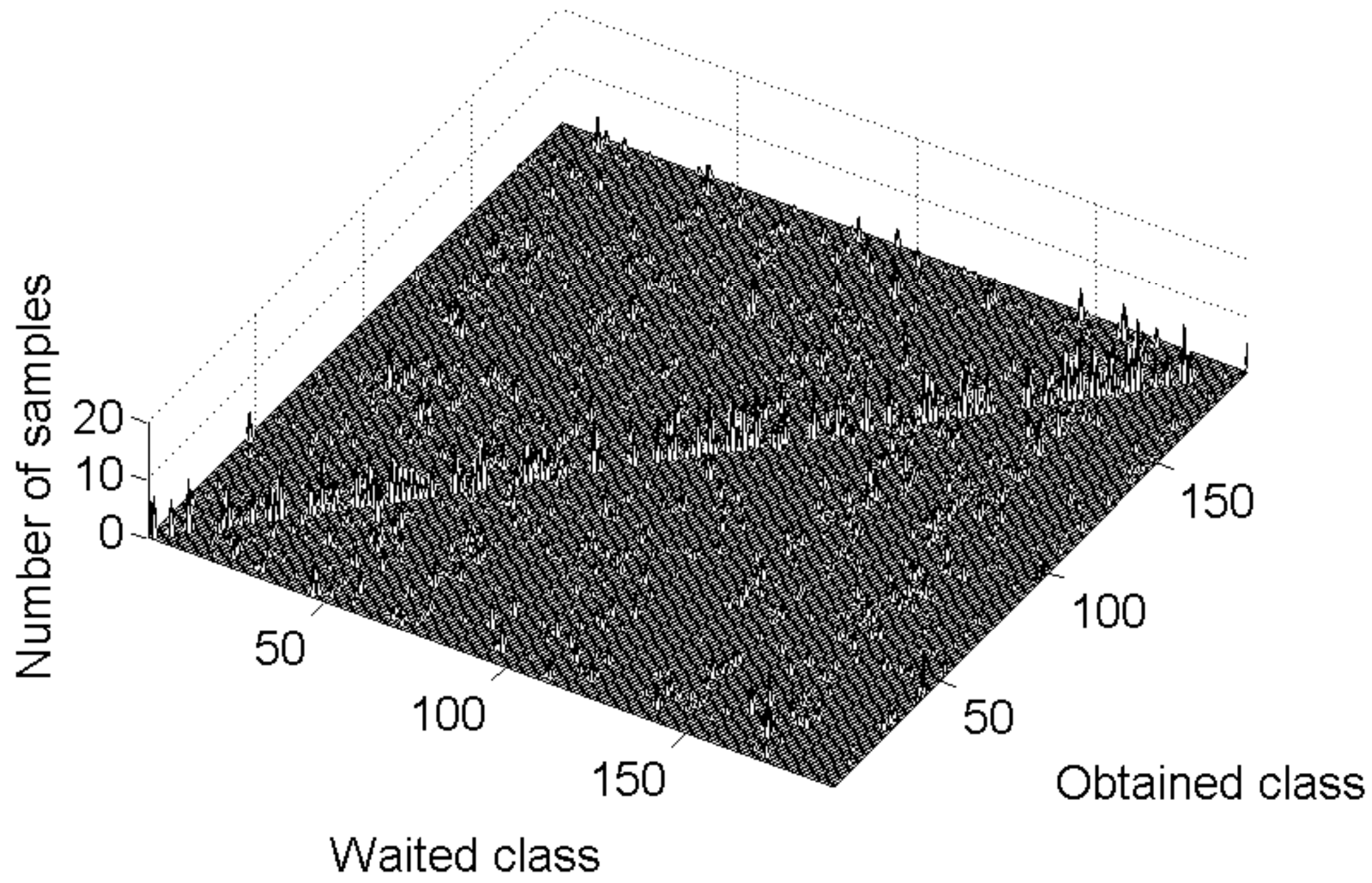}}
								 \subfigure[]{\includegraphics[width=7cm]{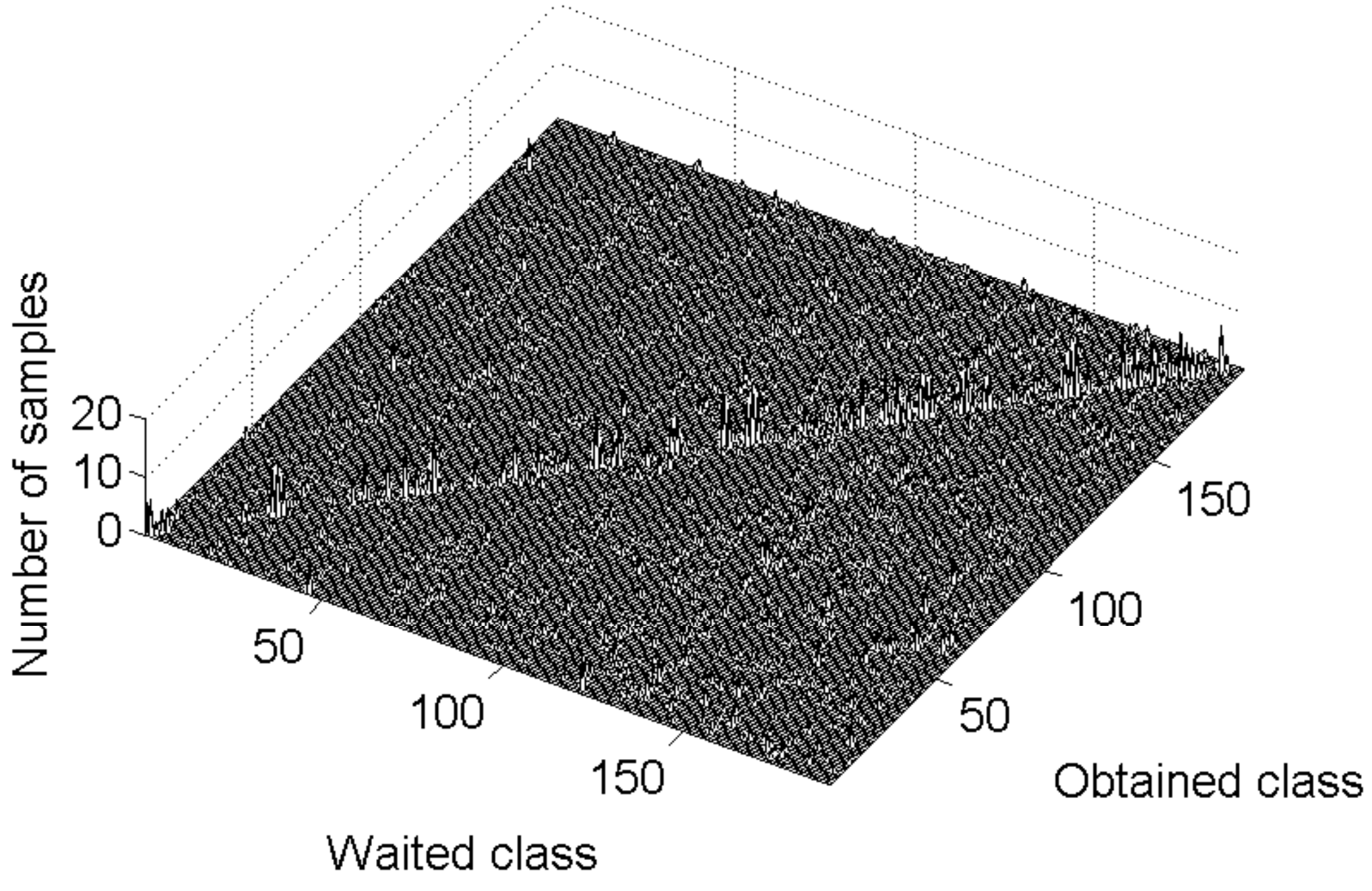}}}					 
					 \mbox{\subfigure[]{\includegraphics[width=7cm]{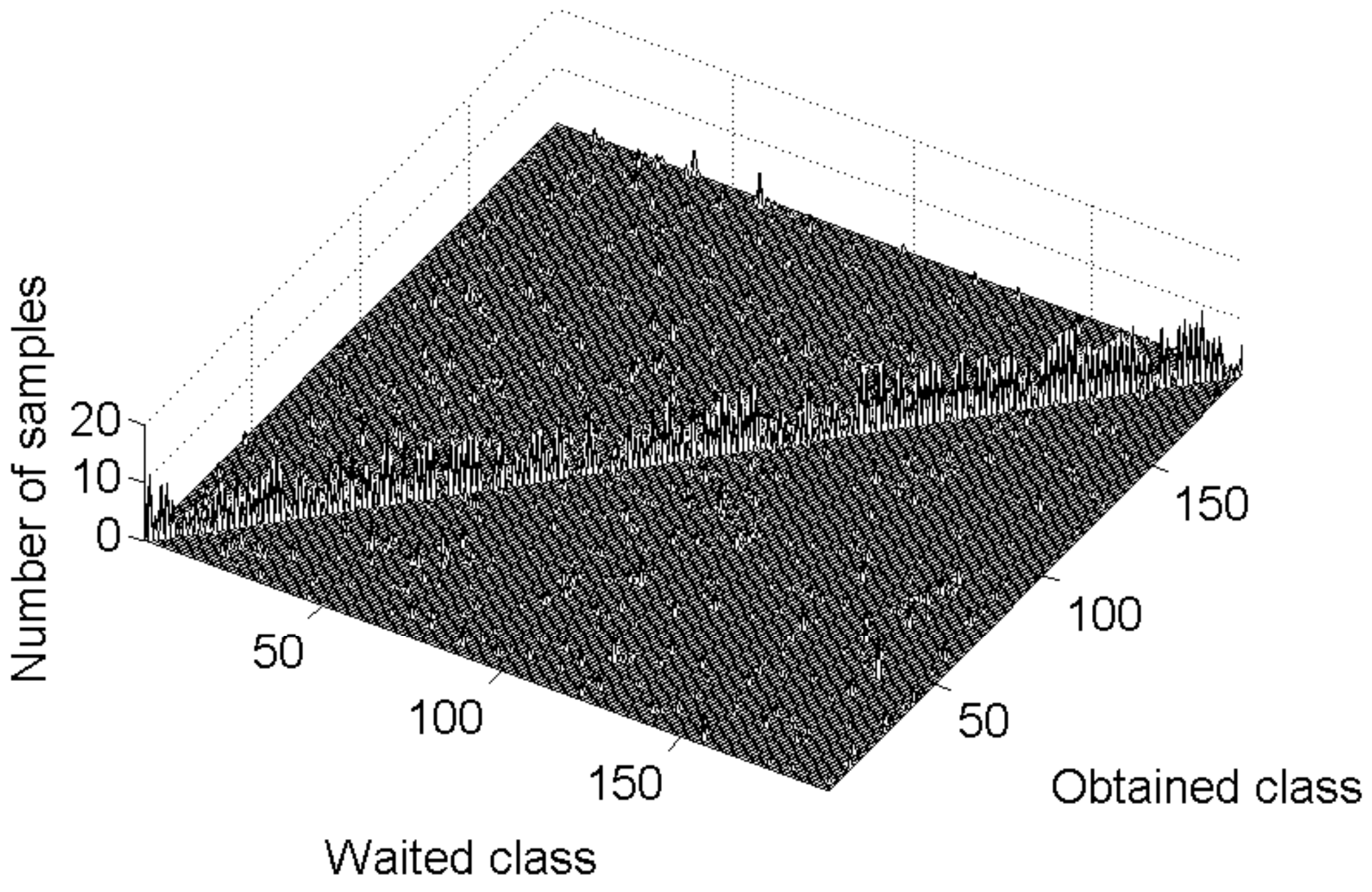}}}					 					 
           \caption{Confusion matrices for each descriptor method in USPTex dataset. Each matrix is mapped onto an image in wich each value in the original image corresponds to a color in the image (see the colorbar legend). a) Box-counting. b) Brownian. c) Bouligand-Minkowski. d) Multifractal. e) Fourier.}
           \label{fig:usptex_conf}                                  
   \end{figure}
\begin{figure}[!htbp] 
					 \centering
           \mbox{\subfigure[]{\includegraphics[width=7cm]{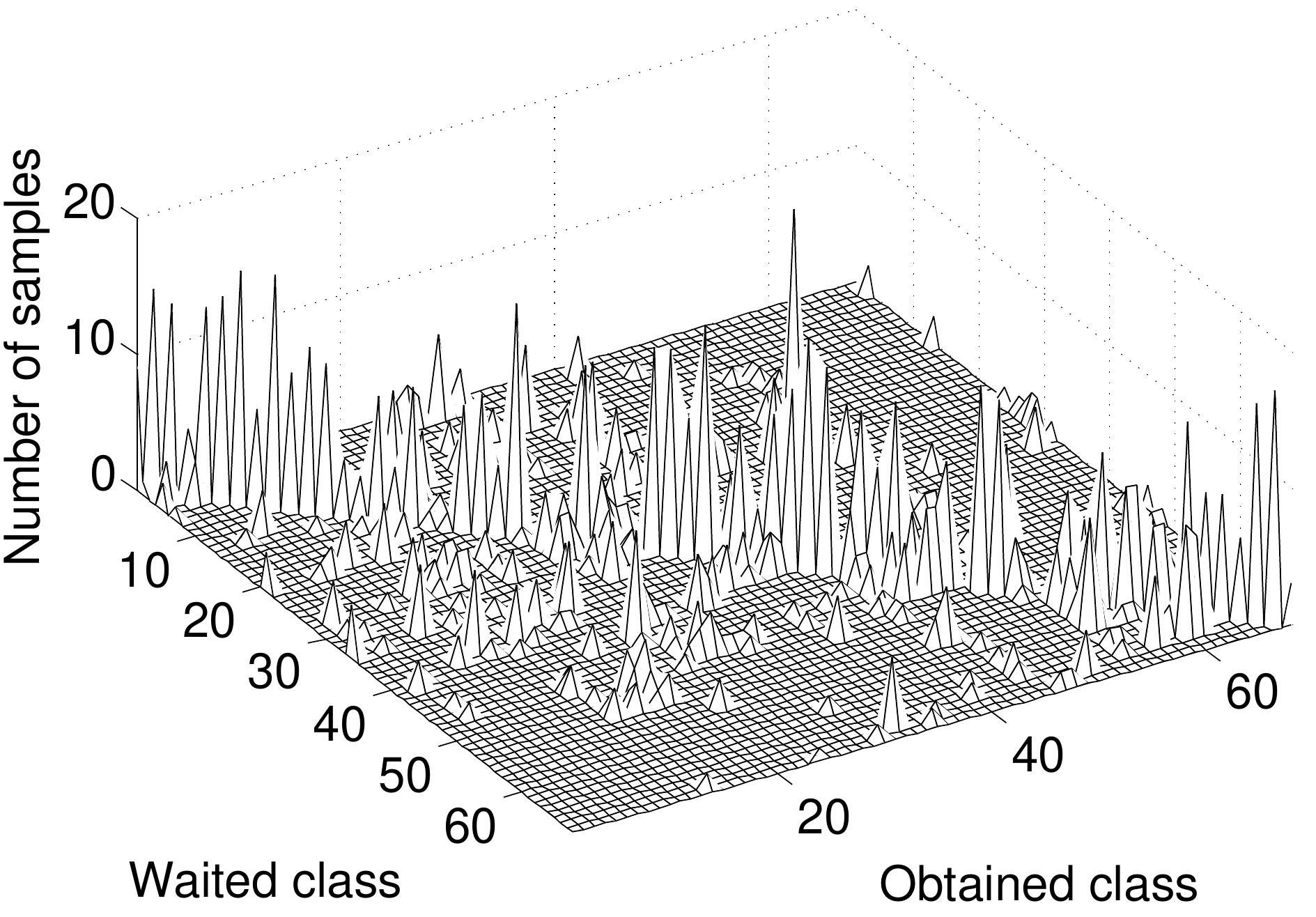}}
           			 \subfigure[]{\includegraphics[width=7cm]{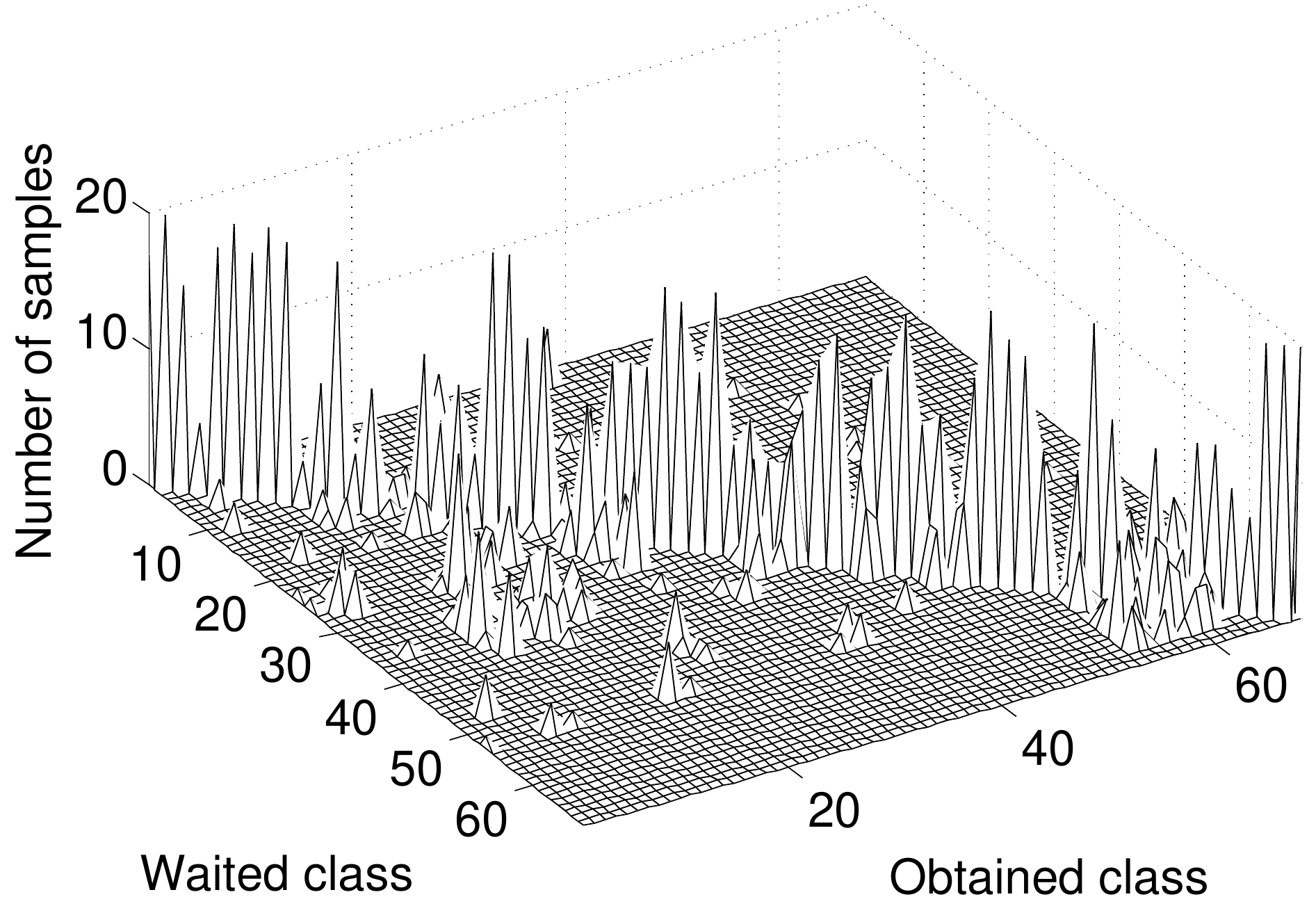}}}
					 \mbox{\subfigure[]{\includegraphics[width=7cm]{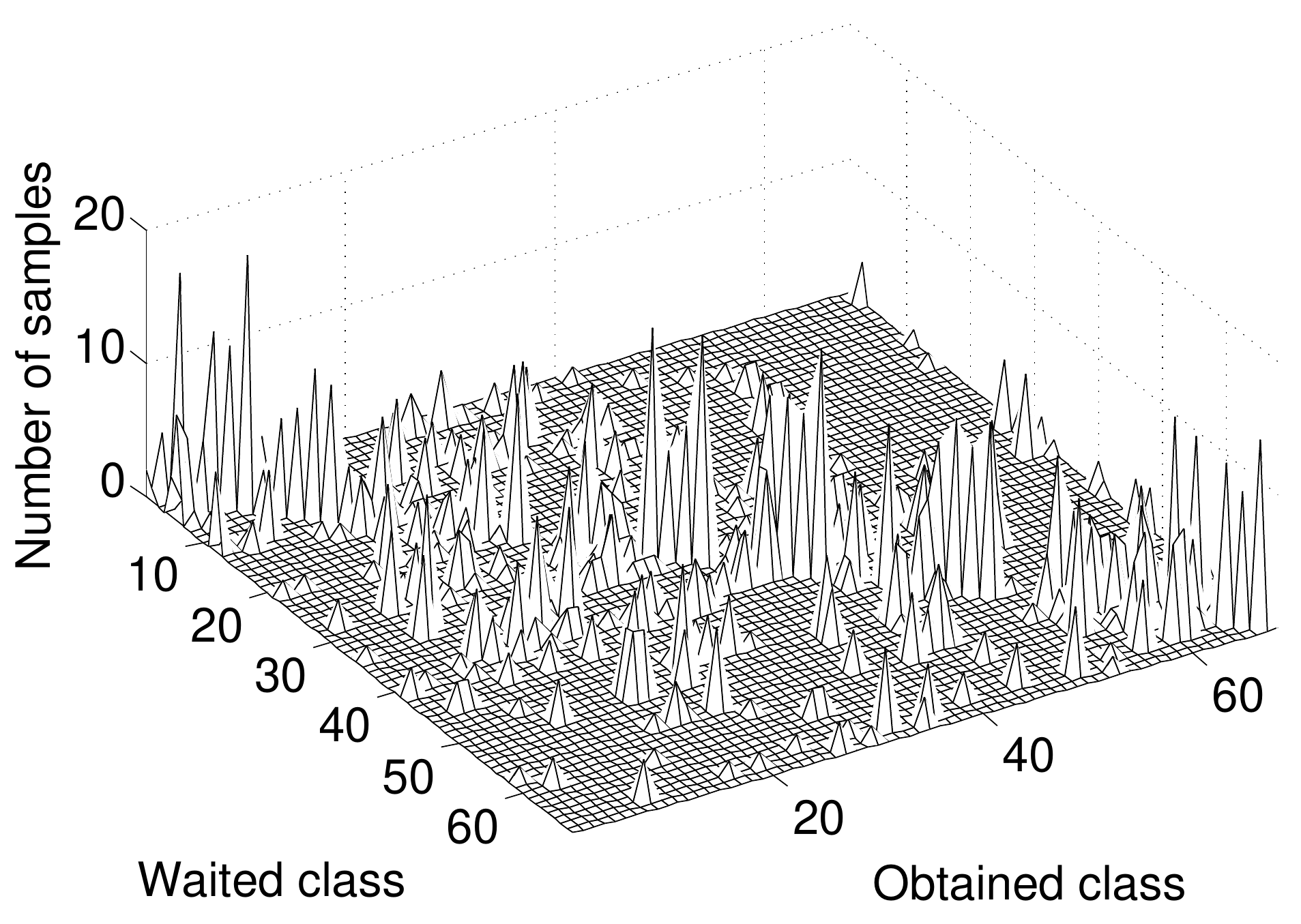}}
								 \subfigure[]{\includegraphics[width=7cm]{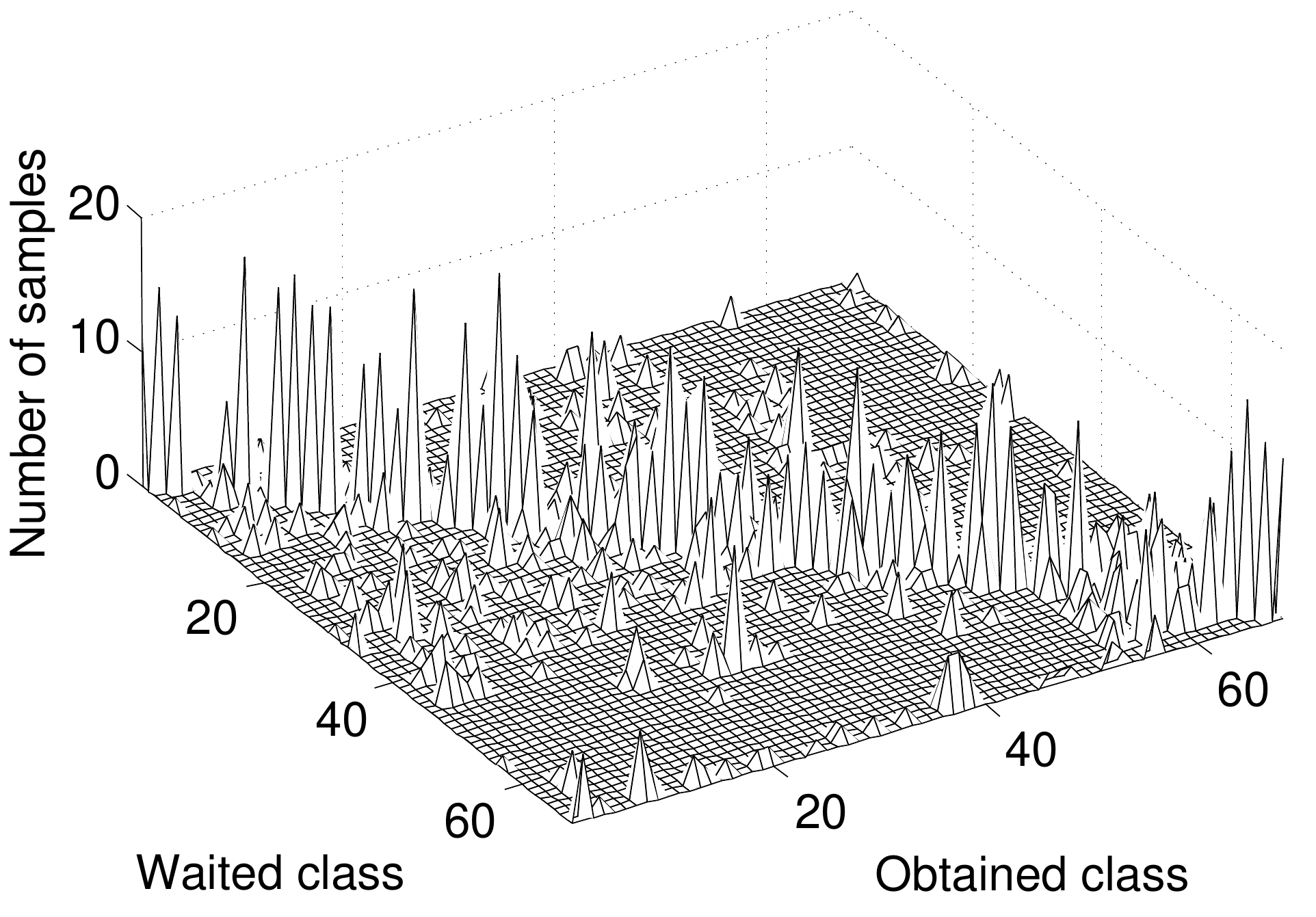}}}					 
					 \mbox{\subfigure[]{\includegraphics[width=7cm]{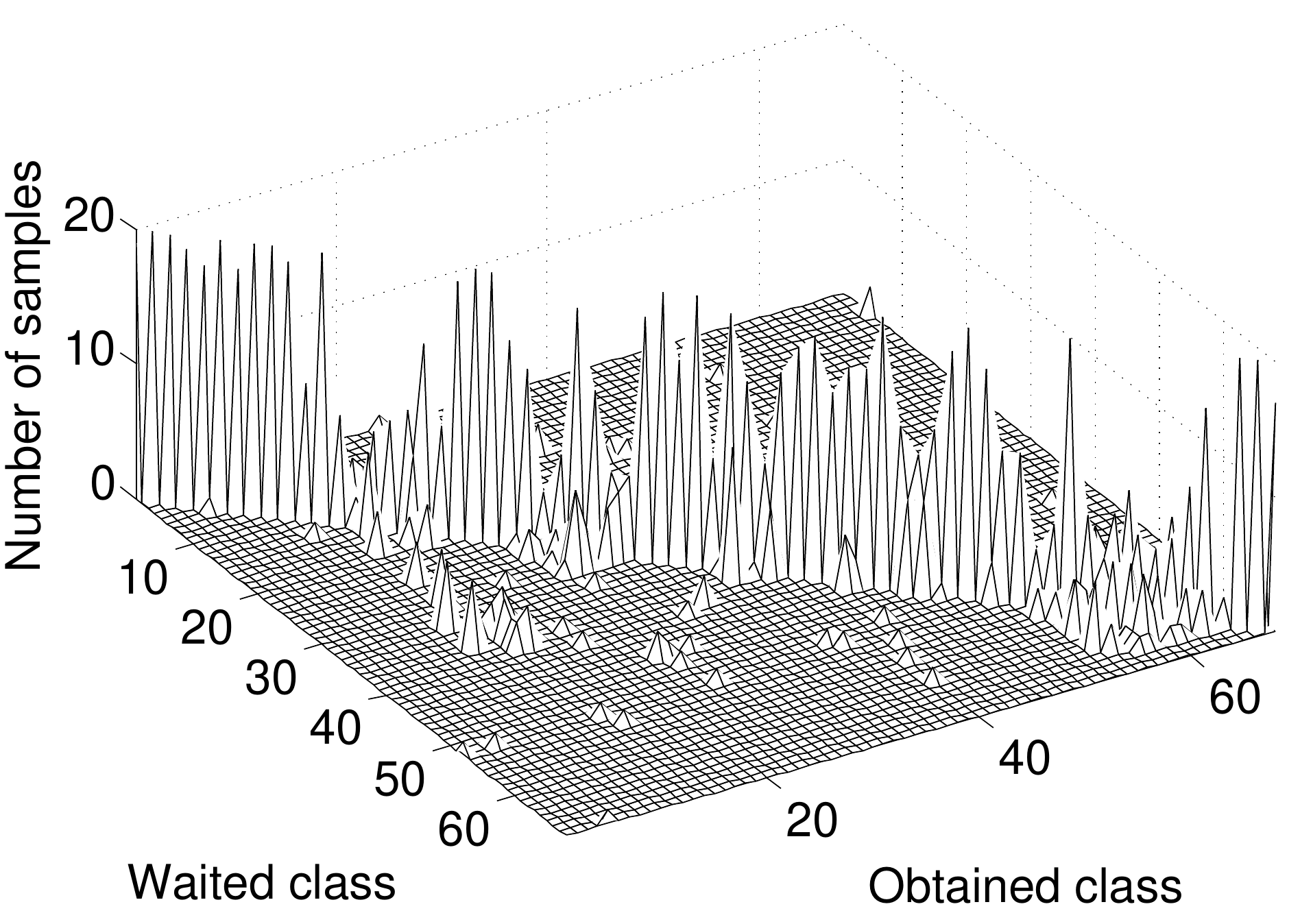}}}								 
           \caption{Confusion matrices for each descriptor method in OuTex dataset. Each matrix is mapped onto an image in wich each value in the original image corresponds to a color in the image (see the colorbar legend). a) Box-counting. b) Brownian. c) Bouligand-Minkowski. d) Multifractal. e) Fourier.}
           \label{fig:outex_conf}                                  
   \end{figure}
\begin{figure}[!htbp] 
					 \centering
           \mbox{\subfigure[]{\includegraphics[width=7cm]{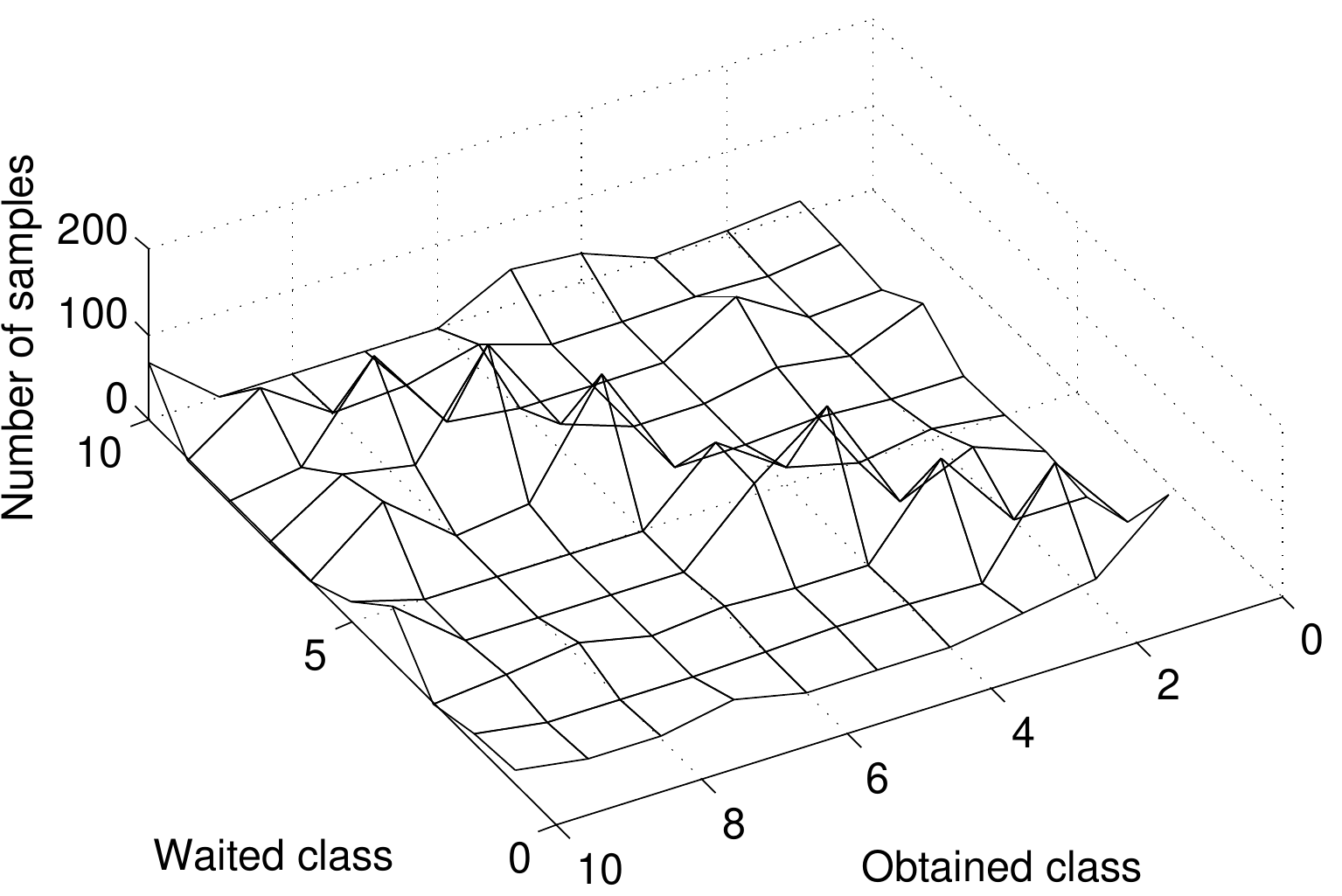}}
           			 \subfigure[]{\includegraphics[width=7cm]{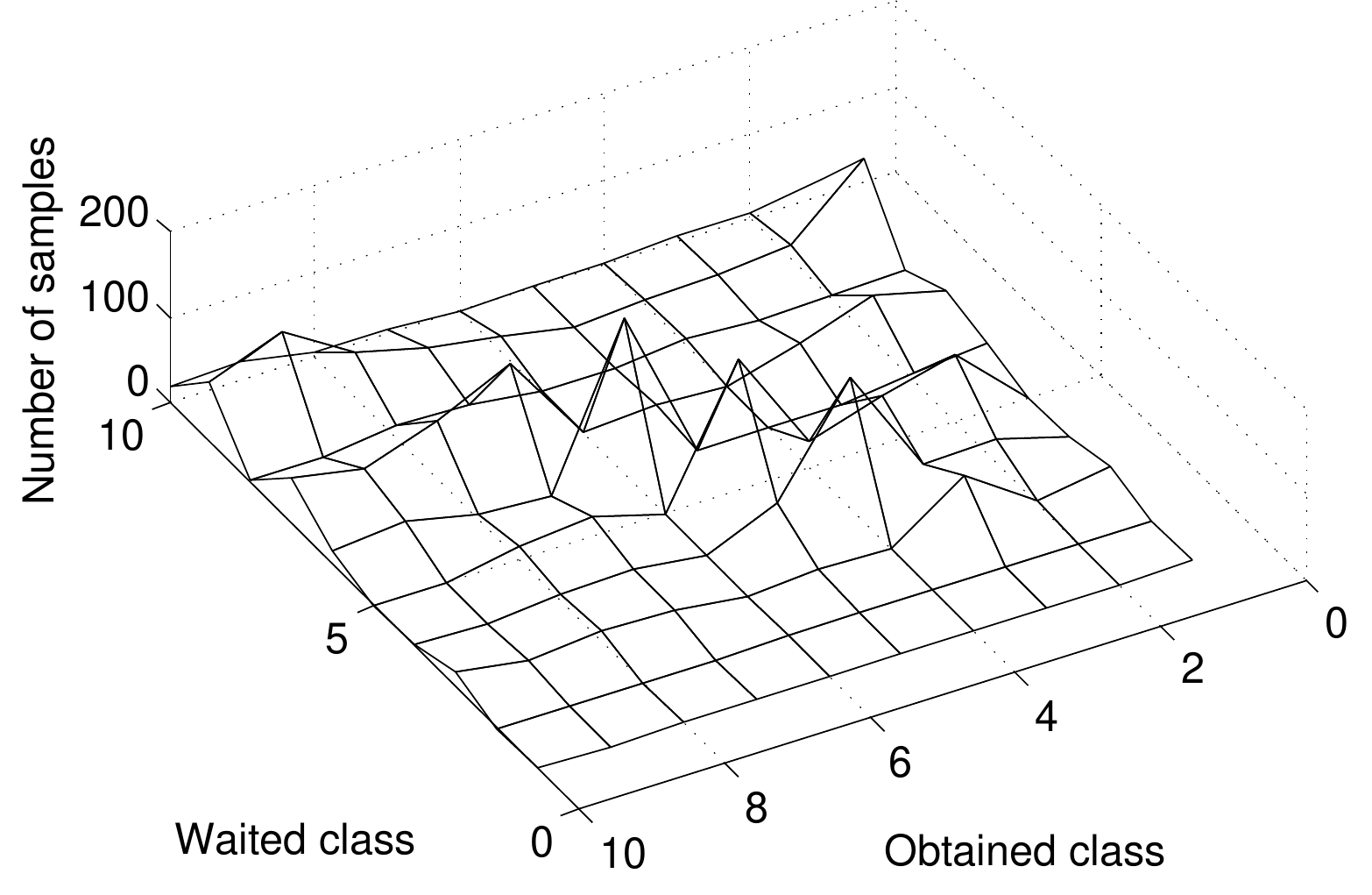}}}
					 \mbox{\subfigure[]{\includegraphics[width=7cm]{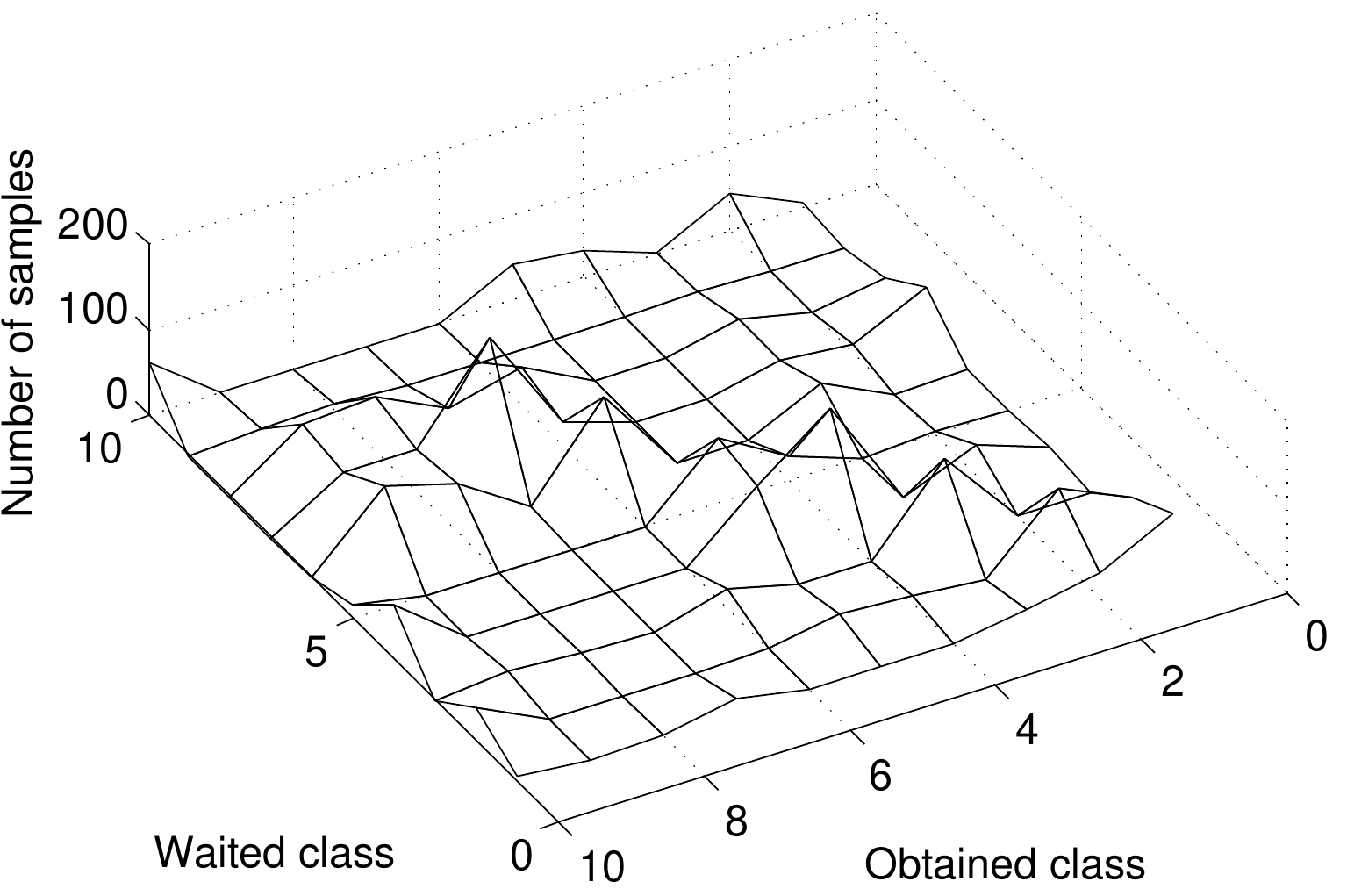}}
								 \subfigure[]{\includegraphics[width=7cm]{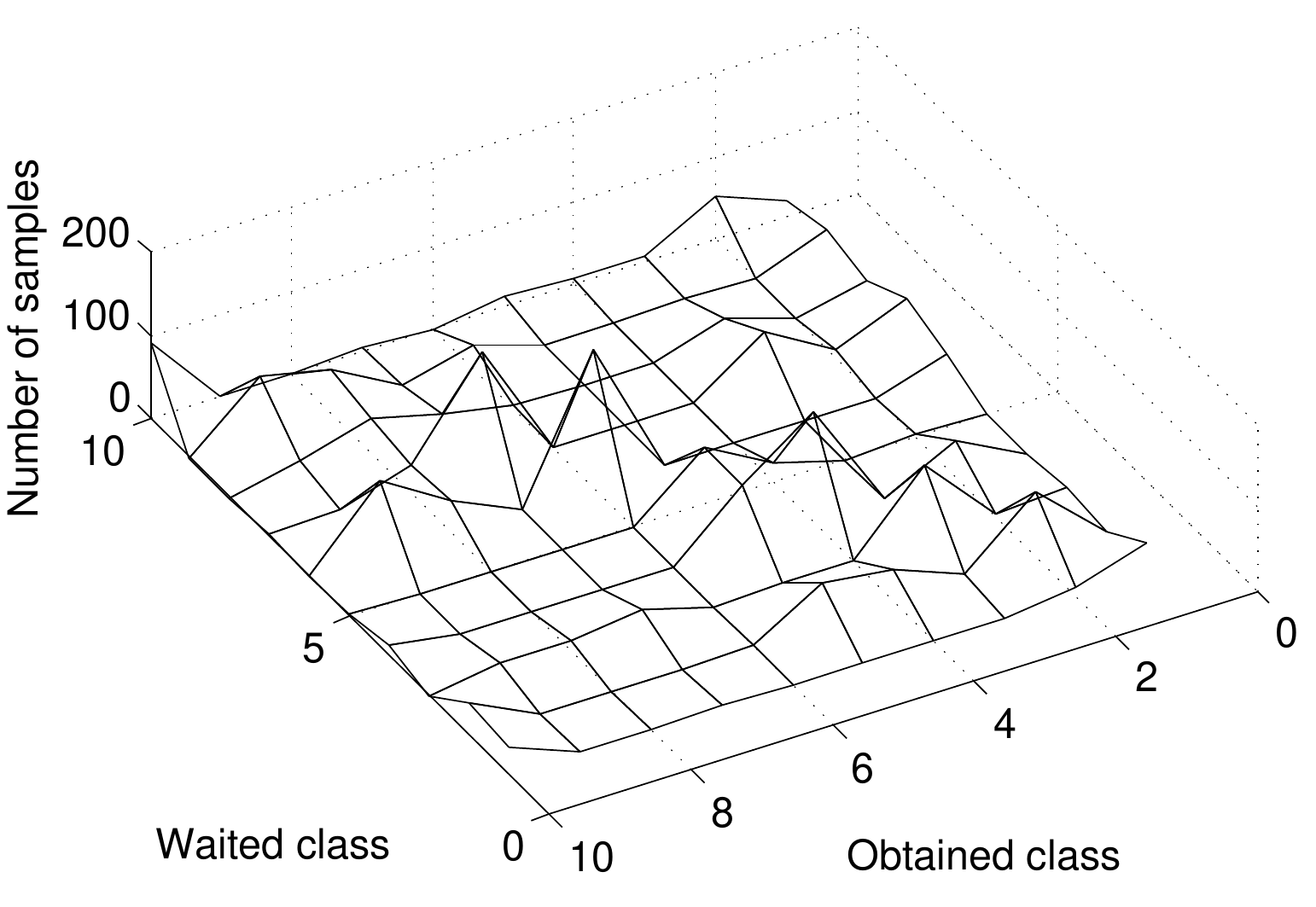}}}					 
					 \mbox{\subfigure[]{\includegraphics[width=7cm]{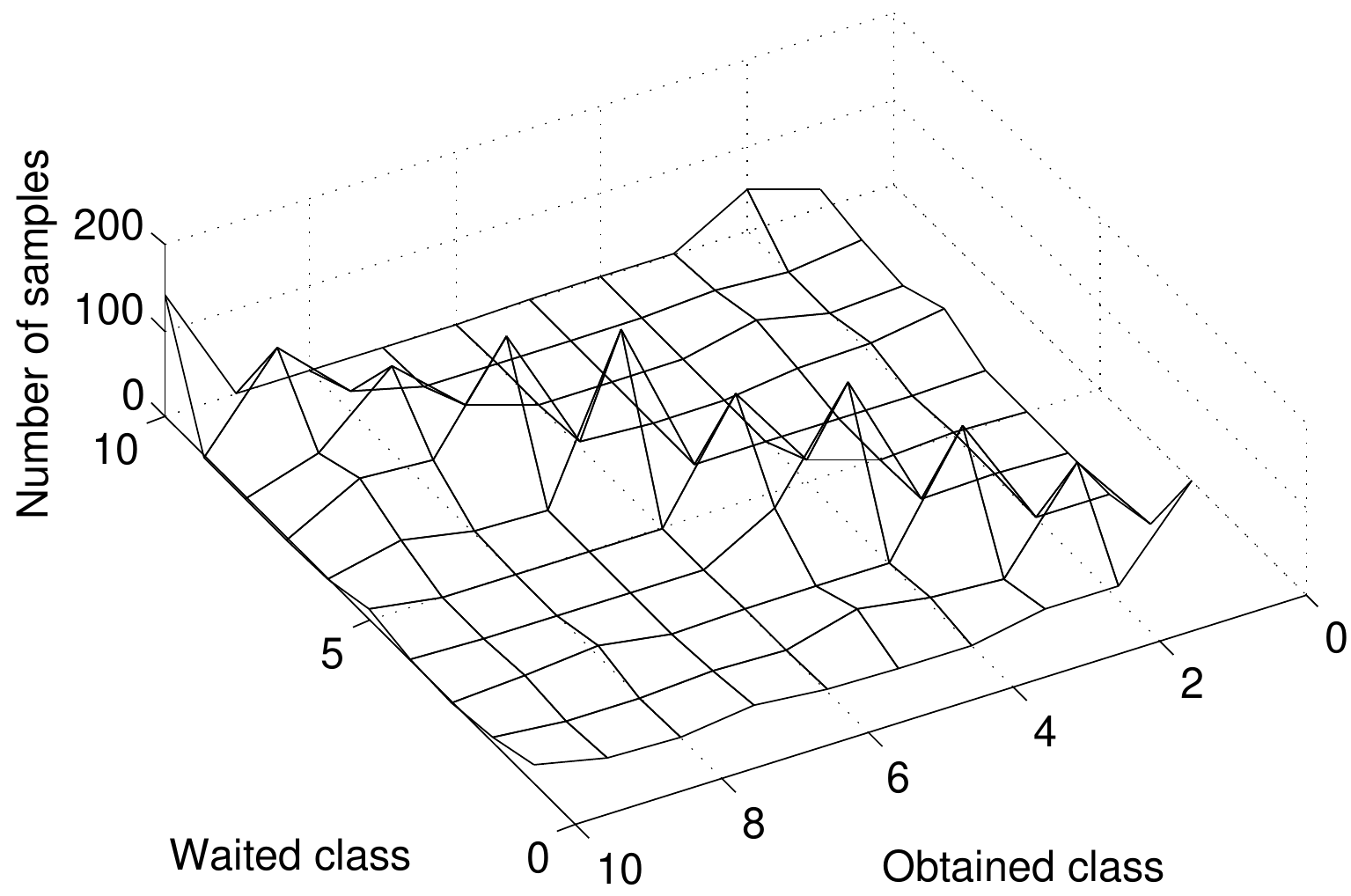}}}								 
           \caption{Confusion matrices for each descriptor method in the leaves dataset. Each matrix is mapped onto a surface in wich each value in the original matrix corresponds to a height in the surface (see the legend). a) Box-counting. b) Brownian. c) Bouligand-Minkowski. d) Multifractal. e) Fourier.}
           \label{fig:folhas_conf}                                  
   \end{figure}

In the Table \ref{tab:brodatz} we observe the best correctness rate achieved by each descriptor in the classification of Brodatz textures. The proposed Fourier fractal descriptor presents the greater correctness. Although it uses more descriptors than other techniques, the number of 74 descriptors does not compromise the computational cost of the classification process. We notice a significant advantage of Fourier descriptors in more than 40\% relative to the second best method, that is, box-counting.
\begin{table*}[!htpb]
	\centering
	\scriptsize
		\begin{tabular}{|c|c|c|}
			\hline
                 Method                        & Correctness Rate (\%) & Number of descriptors\\
                 \hline
Minkowski & 41.3514 & 70\\
Brownian & 56.3063 & 40\\
Box-counting & 52.7027 & 6\\
Multifractal & 35.8559 & 85\\
Fourier & 74.0541 & 74\\			
								 \hline			
		\end{tabular}
	\caption{Correctness rate for the Brodatz dataset.}
	\label{tab:brodatz}
\end{table*}
The Table \ref{tab:usptex} shows the correctness rates for the usptex dataset. The behavior of each method is similar to that found in the Brodatz dataset, although the correctness is smaller for the compared methods, due to the grater number of classes in the dataset.
\begin{table*}[!htpb]
	\centering
	\scriptsize
		\begin{tabular}{|c|c|c|}
			\hline
                 Method                        & Correctness Rate (\%) & Number of descriptors\\
                 \hline
Minkowski & 25.4363 & 25\\
Brownian & 28.1414 & 65\\
Box-counting & 41.6667 & 8\\
Multifractal & 21.3351 & 101\\
Fourier & 54.5375 & 68\\                 
								 \hline			
		\end{tabular}
	\caption{Correctness rate for the USPTex dataset.}
	\label{tab:usptex}
\end{table*}
The Table \ref{tab:outex} shows the global correctness rate for each compared method in the OuTex dataset. One more time, Fourier method presented the greater rate in the classification process. Although Brownian and Box-counting needed fewer descriptors for the best result, as can be seen in the Figure \ref{fig:outex_desc}, the increase in the number of descriptors does not imply in the increase in the correctness rate and the global behavior is inferior to that of the proposed method.
\begin{table*}[!htpb]
	\centering
	\scriptsize
		\begin{tabular}{|c|c|c|}
			\hline
                 Method                        & Correctness Rate (\%) & Number of descriptors\\
                 \hline
Minkowski & 32.2794 & 85\\
Brownian & 58.6029 & 30\\
Box-counting & 41.6176 & 6\\
Multifractal & 40.4412 & 101\\
Fourier & 67.1324 & 68\\                 
								 \hline			
		\end{tabular}
	\caption{Correctness rate for the OuTex dataset.}
	\label{tab:outex}
\end{table*}
Finally, the Table \ref{tab:folhas} shows the global correctness for leaves data set. Fourier have presented a significant advantage of 25.7\% over box-counting descriptors. Again, Fourier method demonstrates to be the best choice in this texture analysis task.
\begin{table*}[!htpb]
	\centering
	\scriptsize
		\begin{tabular}{|c|c|c|}
			\hline
                 Method                        & Correctness Rate (\%) & Number of descriptors\\
                 \hline
Minkowski & 43.2308 & 10\\
Brownian & 40.9744 & 20\\
Box-counting & 54.6667 & 8\\
Multifractal & 46.9744 & 101\\
Fourier & 68.7179 & 44\\                 
								 \hline			
		\end{tabular}
	\caption{Correctness rate for the plant leaves dataset.}
	\label{tab:folhas}
\end{table*}
The Figure \ref{fig:brodatz_conf} shows the confusion matrix for each method in the Brodatz dataset. In the figures, the confusion matrix is depicted in 3D surfaces, in which each point corresponds to a specific column and line in the matrix and the height of the point corresponds to the value in the respective matrix. In these figures, the best methods must present higher regions in the principal diagonal. Besides, the points outside the diagonal must tend to present smaller height. In this sense, the Fourier matrix presents the diagonal with more higher regions and the smaller number of lower points outside de diagonal.
The Figure \ref{fig:usptex_conf} exhibits the confusion matrices fot the USPTex dataset. Again, we observe the diagonal in the Fourier matrix with higher regions.      
The Figure \ref{fig:outex_conf} shows the confusion matrices for the OuTex dataset. Again, the principal diagonal is more continuous with greater heights in the Fourier matrix. We can still observe that the pixels with smaller heights outside the diagonal represent the misclassified samples. In the Fourier matrix, we observe that such lower points are less scattered and are nearer to the diagonal corresponding to a better precision in the discrimination of classes.       
The Figure \ref{fig:folhas_conf} shows the confusion matrices in leaves experiment. One more time, in Fourier descriptors we have the higher regions on the principal diagonal. The other compared techniques present more elevation regions outside the diagonal attesting the presence of more severe misclassifications.
In the Table \ref{tab:time} we present the average computational time (in seconds) expended by each one of the compared techniques in the calculus of descriptors from each image in the analyzed datasets. It is visible the advantage of the proposed Fourier technique. Fourier employed approximately only 14 \% of the time used by the Brownian descriptors, the second faster method.
\begin{table*}[!htpb]
	\centering
	\scriptsize
		\begin{tabular}{|c|c|}
			\hline
                 Method                        & Computational time (seconds)\\
                 \hline
Minkowski & 2.987320\\
Brownian & 0.374237\\
Box-counting & 23.818369\\
Fourier & 0.051799\\                 
								 \hline			
		\end{tabular}
	\caption{Average computational time used by each descriptor technique for each image in the tested datasets.}
	\label{tab:time}
\end{table*}   

\section{Conclusions}

The present work proposed the development and study of a novel texture descriptor based on fractal dimension. Here we proposed the use of Fourier fractal dimension to provide the descriptors. More specifically, the descriptors were obtained from the curve of Fourier power spectrum as a function of the frequency.

The novel descriptors were compared to other approaches in which the descriptors are extracted from the fractal dimension, that is, the Bouligand-Minkowski, the box-counting, brownian motion and multifractal. The comparison was performed over two texture datasets, the Brodatz and the USPTex. The results were favorable to the proposed technique. Fractal Fourier descriptors presented the best correctness rate in the process of classification of textures based on the compared descriptors.

The achieved results suggest is potentially an interesting technique which may be tested in applications relative to texture discrimination, in problems like classification, segmentation and general modeling of textures represented in digital images.

\section{Acknowledgements}
\label{sec:Acknowledgements}
Odemir M. Bruno gratefully acknowledges the financial support of CNPq (National Council for Scientific and Technological Development, Brazil) (Grant \#308449/2010-0 and \#473893/2010-0) and FAPESP (The State of S\~ao Paulo Research Foundation) (Grant \# 2011/01523-1). Jo\~ao B. Florindo is grateful to CNPq(National Council for Scientific and Technological Development, Brazil) for his doctorate grant.

\newpage


\end{document}